\documentclass[a4paper, 11pt]{article}
\pdfoutput=1
\usepackage{jheppub}

\usepackage{cancel}
\usepackage{verbatim} 
\newcommand{\al}{\ensuremath{\alpha} }
\newcommand{\be}{\ensuremath{\beta} }
\newcommand{\ga}{\ensuremath{\gamma} }
\newcommand{\de}{\ensuremath{\delta} }
\newcommand{\gaEff}{\ensuremath{\ga_{\rm eff}} }
\newcommand{\si}{\ensuremath{\sigma} }
\newcommand{\cN}{\ensuremath{\mathcal N} }
\newcommand{\cO}{\ensuremath{\mathcal O} }
\newcommand{\gc}{\ensuremath{g_c^2} }
\newcommand{\gGF}{\ensuremath{g_{\rm GF}^2} }
\newcommand{\gtc}{\ensuremath{\widetilde g_c^2} }
\newcommand{\gtGF}{\ensuremath{\widetilde g_{\rm GF}^2} }
\newcommand{\topt}{\ensuremath{\tau_{\rm opt}} }
\newcommand{\gsim}{\ensuremath{\gtrsim} }
\newcommand{\lsim}{\ensuremath{\lesssim} }
\newcommand{\X}{\ensuremath{\!\times\!} }
\newcommand{\Sb}{\ensuremath{\cancel{S^4}} }
\newcommand{\MSbar}{\ensuremath{\overline{\textrm{MS}} } }
\newcommand{\chidof}{\ensuremath{\mbox{$\chi^2/\text{d.o.f.}$}}}
\newcommand{\vev}[1]{\ensuremath{\left\langle #1 \right\rangle} }
\newcommand{\nn}{\nonumber}
\newcommand{\eq}[1]{eq.~\ref{#1}}
\newcommand{\fig}[1]{figure~\ref{#1}}
\newcommand{\Fig}[1]{Figure~\ref{#1}}
\newcommand{\tab}[1]{table~\ref{#1}}
\newcommand{\secref}[1]{section~\ref{#1}}
\newcommand{\refcite}[1]{ref.~\cite{#1}}

\title{Nonperturbative \be function \\ of eight-flavor SU(3) gauge theory}

\author{Anna~Hasenfratz,$^1$}
\author{David~Schaich$^2$}
\author{and Aarti~Veernala$^2$}
\affiliation{$^1$Department of Physics, University of Colorado, Boulder, Colorado 80309, U.S.A.}
\affiliation{$^2$Department of Physics, Syracuse University, Syracuse, New York 13244, U.S.A.}
\emailAdd{anna.hasenfratz@colorado.edu}
\emailAdd{dschaich@syr.edu}
\emailAdd{aveernal@syr.edu}

\abstract{ 
  We present a new lattice study of the discrete \be function for SU(3) gauge theory with $N_f = 8$ massless flavors of fermions in the fundamental representation.
  Using the gradient flow running coupling, and comparing two different nHYP-smeared staggered lattice actions, we calculate the 8-flavor step-scaling function at significantly stronger couplings than were previously accessible.
  Our continuum-extrapolated results for the discrete \be function show no sign of an IR fixed point up to couplings of $g^2 \approx 14$.
  At the same time, we find that the gradient flow coupling runs much more slowly than predicted by two-loop perturbation theory, reinforcing previous indications that the 8-flavor system possesses nontrivial strongly coupled IR dynamics with relevance to BSM phenomenology.
}
\arxivnumber{1410.5886}

\begin{document}
\maketitle
\flushbottom

\section{\label{sec:intro}Introduction and review of previous work} 
SU(3) gauge theory with $N_f = 8$ flavors of massless fermions in the fundamental representation is interesting both theoretically and in the context of phenomenology for new physics beyond the standard model (BSM).
Theoretical interest comes from the possibility that $N_f = 8$ may be close to the lower boundary of the conformal window, the range of $N_f^{(c)} \leq N_f < 16.5$ for which the theory flows to a chirally symmetric conformal fixed point in the infrared (IRFP)~\cite{Caswell:1974gg, Banks:1981nn}.
The connection to BSM phenomenology stems from expectations that mass-deformed models with $N_f$ near $N_f^{(c)}$ will possess strongly coupled approximately conformal dynamics, producing a large mass anomalous dimension and slowly running (``walking'') gauge coupling across a wide range of energy scales~\cite{Holdom:1981rm, Yamawaki:1985zg, Appelquist:1986an}.
In models of new strong dynamics based on 8-flavor SU(3) gauge theory, these features are invoked to evade constraints from flavor-changing neutral currents, to produce a phenomenologically viable electroweak $S$ parameter, and to justify a relatively light and SM-like composite Higgs boson with mass $M_H = 125$~GeV.
See refs.~\cite{Appelquist:2013sia, Kuti:2014epa} for brief reviews of these issues.

The onset of IR conformality is an inherently nonperturbative phenomenon: at the two-loop perturbative level, the conformal window opens with the appearance of an IR fixed point in the infinite-coupling limit.
This occurs at a non-integer $N_f^{(c)} \approx 8.05$ very close to $N_f = 8$.
Although both three- and four-loop perturbative calculations of the renormalization group \be function in the \MSbar scheme predict an 8-flavor IRFP, the resulting fixed-point coupling is still quite strong, $g_{\MSbar}^2 \approx 18.4$ and 19.5, respectively~\cite{Ryttov:2010iz}.
There is no reason to trust perturbation theory at such strong couplings.
We also do not wish to rely on arguments that spontaneous chiral symmetry breaking should be induced for $g_{\MSbar}^2 \sim 10$, which combine perturbation theory with an approximate analysis of Schwinger--Dyson equations~\cite{Appelquist:1998rb}. 
The resulting estimates of $N_f^{(c)}$ based on Schwinger--Dyson equations range from $N_f^{(c)} \approx 8$ in \refcite{Bashir:2013zha} to $N_f^{(c)} \approx 12$ in \refcite{Appelquist:1996dq}, while a bound $N_f^{(c)} \lsim 12$ follows from a conjectured thermal inequality~\cite{Appelquist:1999hr}.

Since interest in 8-flavor SU(3) gauge theory revolves around its strongly coupled IR dynamics, lattice gauge theory is an indispensable approach to study the system nonperturbatively, from first principles.
A wide variety of methods have been employed by existing lattice studies.
These include investigation of the running coupling and its discrete \be function~\cite{Appelquist:2007hu, Appelquist:2009ty}; exploration of the phase diagram through calculations at finite temperature~\cite{Deuzeman:2008sc, Miura:2012zqa, Jin:2010vm, Schaich:2012fr, Hasenfratz:2013uha}; analysis of hadron masses and decay constants~\cite{Fodor:2009wk, Aoki:2013xza, Aoki:2014oha, Schaich:2013eba, Appelquist:2014zsa}; and study of the eigenmodes of the Dirac operator~\cite{Fodor:2009wk, Cheng:2013eu}.
These various analyses are complementary, and in combination offer the most reliable information about the IR dynamics of the system.
Let us summarize the strengths of each of these approaches, and review the current state of knowledge for the 8-flavor system, to motivate the new work that we will present.

In this paper we will report on a new step-scaling study of the 8-flavor discrete \be function, exploiting several recent improvements to this method.
Generically, running coupling studies are carried out in the $am = 0$ chiral limit, and connect the perturbative (asymptotically free) UV regime to the strongly coupled IR.
The IR limit of a massless theory is characterized by either spontaneous chiral symmetry breaking or renormalization group flow to an IR fixed point.
Lattice running coupling studies, after extrapolation to the continuum, directly search for an IRFP within the range of renormalized couplings probed by the study.
At the same time, the use of massless fermions prevents these studies from exploring chirally broken dynamics, which finite-temperature or spectral techniques are better suited to investigate.

For example, the (pseudo)critical couplings $g_{cr}$ of chiral transitions at finite temperature $T$ and nonzero fermion mass $am$ depend on the lattice spacing $a$, or equivalently on the temporal extent of the lattice $N_t = 1 / (aT)$.
In a chirally broken system, these transitions must move to the asymptotically free UV fixed point $g_{cr} \to 0$ as the UV cutoff $a^{-1} \to \infty$.
At the same time the fermion mass must be extrapolated to the $am \to 0$ chiral limit to ensure that the observed chiral symmetry breaking is truly spontaneous.
In an IR-conformal system, in contrast, the finite-temperature transitions in the chiral limit must accumulate at a finite coupling as $N_t \to \infty$, and remain separated from the weak-coupling conformal phase by a bulk transition.
Spectral studies can proceed more directly by attempting to fit nonzero-mass lattice data to chiral perturbation theory.
Since the chiral regime is inaccessible to existing studies, these investigations typically search for simpler signs that the pseudoscalar mesons behave as Goldstone bosons in the chiral limit, for instance by considering whether the ratio of vector and pseudoscalar meson masses $M_V / M_P \to \infty$ as $am \to 0$.
In a similar vein, eigenmode studies can investigate chiral symmetry breaking by comparing the low-lying Dirac spectrum with random matrix theory, or by considering the scale dependence of the effective mass anomalous dimension predicted by the eigenmode number.

Spectral and eigenmode studies have further applications beyond simply searching for spontaneous chiral symmetry breaking.
The hadron masses themselves are phenomenologically interesting.
In addition to exploring whether the system may possess a sufficiently light Higgs particle, these calculations predict the properties of further resonances that may be observed at the Large Hadron Collider or future experiments.
The low-energy constants of the effective chiral Lagrangian are also experimentally accessible, for example in the form of the electroweak $S$ parameter and WW scattering lengths~\cite{Appelquist:2010xv, Appelquist:2012sm}.
Finally, in approximately conformal systems, finite-size scaling of spectral data can probe the effective mass anomalous dimension $\gaEff(\mu)$, the scale dependence of which can be extracted from the Dirac eigenmodes~\cite{Cheng:2013eu, Cheng:2013bca}.

In the context of the 8-flavor system, a pioneering lattice investigation performed a running coupling study based on the Schr\"odinger functional~\cite{Appelquist:2007hu, Appelquist:2009ty}.
This work could access the continuum-extrapolated discrete \be function up to $g_{SF}^2 \lesssim 6.6$, in which range reasonable agreement with two-loop perturbation theory was found. 
In part, computational expense limited the strength of the renormalized coupling that could be considered.
In addition, the study had to avoid a bulk phase transition at stronger bare couplings, a typical restriction that prevents lattice calculations from probing arbitrarily strong couplings.
Since refs.~\cite{Appelquist:2007hu, Appelquist:2009ty} used unimproved staggered fermions, one may expect to reach stronger couplings and to reduce computational costs by improving the lattice action, which is one of the steps we take in the present work.

Given the evidence from refs.~\cite{Appelquist:2007hu, Appelquist:2009ty} for rough consistency with perturbation theory up to $g_{SF}^2 \approx 6.6$, we can turn to finite-temperature and spectral studies to explore whether chiral symmetry is spontaneously broken at these couplings.
The pioneering 8-flavor finite-temperature study of \refcite{Deuzeman:2008sc}, later extended by \refcite{Miura:2012zqa}, investigated $N_t = 6$, 8 and 12 with fixed $am = 0.02$,\footnote{In this discussion we don't attempt to compare the lattice scales $a^{-1}$ between different studies.  As a consequence, a given numerical value for $am$ will correspond to slightly different physical masses for different lattice actions.  However, the broad consistency between these studies implies that they are at least qualitatively comparable.} for which mass the chiral transitions move to weaker coupling for larger $N_t$ in agreement with two-loop perturbation theory. 
In order to explore the approach to the chiral limit, in recent work we carried out finite-temperature investigations for a range of fermion masses $am \leq 0.02$ with $N_t = 12$, 16 and 20~\cite{Schaich:2012fr, Hasenfratz:2013uha}.
For sufficiently large masses $am \geq 0.01$ we also observed two-loop scaling, but this did not persist at smaller $am \leq 0.005$, where the finite-temperature transitions merged with a bulk transition into a lattice phase.
(We will review our lattice phase diagram in \secref{sec:setup}.)
Even ongoing studies using a rather large $48^3\X 24$ lattice volume, part of a joint project with the Lattice Strong Dynamics Collaboration, have not yet established spontaneous chiral symmetry breaking, as we will report in a future publication~\cite{LSDfiniteT}.

Similarly, studies of the 8-flavor spectrum and Dirac eigenmodes have not clearly demonstrated spontaneous chiral symmetry breaking.
In \refcite{Aoki:2013xza} the LatKMI Collaboration argued that at lighter fermion masses $0.015 \leq am \leq 0.04$ the spectrum of the theory may be described by chiral perturbation theory, while data at heavier $0.05 \leq am \leq 0.16$ appear to exhibit some remnant of IR conformality despite chiral symmetry breaking.
At smaller masses $0.004 \leq am \leq 0.01$ and larger lattice volumes up to $48^3\X96$, however, a USBSM project could not confirm spontaneous chiral symmetry breaking~\cite{Schaich:2013eba}.
Recent work by the Lattice Strong Dynamics Collaboration using the domain wall fermion formulation (as opposed to the staggered fermions used by all other studies discussed above) observed a slight but steady increase in the ratio $M_V / M_P$ for smaller fermion masses in the range $0.0127 \leq am \leq 0.0327$, even though their data were not within the radius of convergence of chiral perturbation theory.

In summary, although existing lattice studies are all consistent with 8-flavor SU(3) gauge theory being chirally broken, with no evidence for IR conformality, spontaneous chiral symmetry breaking has not yet been conclusively established.
The implications of this situation extend well beyond a simple categorization of the system.
In particular, the lattice results provide indications that the $N_f = 8$ model exhibits the desirable phenomenological features expected for $N_f \approx N_f^{(c)}$.
When analyzed within the framework of mass-deformed IR conformality, the spectral studies mentioned above prefer a large effective mass anomalous dimension $\gaEff \simeq 1$.
Our investigations of Dirac eigenmode scaling find that this large $\gaEff(\mu)$ persists across a wide range of energy scales~\cite{Cheng:2013eu}.\footnote{That is, the 8-flavor system appears to satisfy this condition of ``walking'' dynamics, suggesting that models of new strong dynamics based on this gauge theory may be phenomenologically viable.  The specific role of the scheme- and scale-dependent $\gaEff(\mu)$ within such models is non-trivial to analyze and beyond the scope of this work.}
Arguably the most exciting recent development is the observation of a light flavor-singlet scalar Higgs particle by the LatKMI Collaboration~\cite{Aoki:2014oha}.\footnote{A similar light scalar was also found for $N_f = 12$~\cite{Aoki:2013zsa, Fodor:2014pqa}, a system for which our previous running coupling study observed an IR fixed point in the continuum chiral limit~\cite{Cheng:2014jba}.  Further information about the 12-flavor theory and other potentially interesting systems can be found in the reviews~\cite{Appelquist:2013sia, Kuti:2014epa} and references therein.} 

From these considerations, we conclude that further lattice studies of the 8-flavor system are well motivated.
In this paper we present a new study of the discrete \be function, taking two novel steps in order to access stronger couplings than were previously probed for $N_f = 8$.
First, instead of using the traditional Schr\"odinger functional running coupling discussed above, we employ a recently introduced alternative based on the gradient flow, which offers improved statistical precision for lower computational costs.
We review gradient flow step scaling in the next section, also summarizing several recent improvements that make this method more robust against systematic errors.
In addition, we make use of highly improved lattice actions, comparing two staggered-fermion actions with either one or two nHYP smearing steps.
(The once-smeared action is also being used in separate finite-temperature~\cite{LSDfiniteT}, spectral~\cite{Schaich:2013eba} and eigenmode~\cite{Cheng:2013eu} studies, which offer complementary insight into additional aspects of this system.)
In \secref{sec:setup} we describe our numerical setup and lattice ensembles, focusing on the issue of how to reach strong renormalized couplings without encountering a bulk transition into a lattice phase.

Our step-scaling analyses and results are presented in \secref{sec:results}.
Our nonperturbative study predicts the continuum-extrapolated discrete \be function of 8-flavor SU(3) gauge theory up to renormalized couplings $\gc \approx 14$.
For much of this range we find that the coupling runs much more slowly than in two-loop perturbation theory, and also more slowly than the (IR-conformal) four-loop \MSbar prediction.
We conclude in \secref{sec:conclusion} with discussion of important directions to pursue in further future studies of eight flavors on the lattice.

\section{\label{sec:gradflow}Gradient flow step scaling and its improvement} 
The gradient flow is a continuous transformation that smooths lattice gauge fields to systematically remove short-distance lattice cutoff effects~\cite{Narayanan:2006rf}.
Following the demonstration that the gradient flow is mathematically well defined and invertible~\cite{Luscher:2009eq}, it has been used in a wide variety of applications (recently reviewed by \refcite{Luscher:2013vga}).
We are interested in step-scaling studies of a renormalized coupling defined through the gradient flow.
This coupling is based on the energy density $E(t) = -\frac{1}{2}\mbox{ReTr}\left[G_{\mu\nu}(t) G^{\mu\nu}(t)\right]$ after flow time $t$, which defines~\cite{Luscher:2010iy}
\begin{equation}
  \gGF(\mu) = \frac{1}{\cN} \vev{t^2 E(t)} = \frac{128\pi^2}{3(N^2 - 1)} \vev{t^2 E(t)}
\end{equation}
at energy scale $\mu = 1 / \sqrt{8t}$.
The normalization \cN is set by requiring that $\gGF(\mu)$ agrees with the continuum \MSbar coupling at tree level.
To use the gradient flow coupling in step-scaling analyses, we tie the energy scale to the lattice volume $L^4$ by fixing the ratio $c = \sqrt{8t} / L$, as proposed by Refs.~\cite{Fodor:2012td, Fodor:2012qh, Fritzsch:2013je}.
Each choice of $c$ defines a different renormalization scheme, producing a different renormalized coupling $\gc(L)$ and predicting a different discrete \be function in the continuum limit.
If periodic boundary conditions (BCs) are used for the gauge fields, these \be functions are only one-loop (and not two-loop) universal~\cite{Fodor:2012td}.

At nonzero bare coupling $g_0^2$, the gradient flow renormalized couplings \gc have cutoff effects that must be removed by extrapolating to the $(a / L) \to 0$ continuum limit.
The cutoff effects depend on the lattice action used to generate the configurations, on the gauge action used in the gradient flow transformation, and on the lattice operator used to define the energy density $E(t)$.
While it is possible to systematically remove lattice artifacts by improving all three quantities simultaneously, this approach is not always reasonable in practice.
Another option proposed by \refcite{Fodor:2014cpa} is to modify the definition of the renormalized coupling to perturbatively correct for cutoff effects, 
\begin{equation}
  \label{eq:pert_g2}
  \gc(L) = \frac{128\pi^2}{3(N^2 - 1)} \frac{1}{C(L, c)} \vev{t^2 E(t)}.
\end{equation}
Here the function $C(L, c)$ is a four-dimensional finite-volume sum in lattice perturbation theory, which depends on the action, flow and operator.
It is computed at tree level by \refcite{Fodor:2014cpa}, and we use that result to include this correction in our definition of $\gc$.
Since we use periodic BCs for the gauge fields, the correction $C(L, c)$ also includes a term that accounts for the zero-mode contributions.

Even with this tree-level improvement, the gradient flow step scaling can show significant cutoff effects.
These can be reduced to some extent by working with relatively large $c \gsim 0.3$, at the price of increased statistical uncertainties~\cite{Fritzsch:2013je}.
In \refcite{Cheng:2014jba} we introduced a different modification of the renormalized coupling that replaces the energy density $\vev{E(t)}$ with the value resulting from a small shift in the flow time,
\begin{equation}
  \label{eq:t-shift}
  \gtGF(\mu; a) = \gGF(\mu; a) \frac{\vev{E(t + \tau_0 a^2)}}{\vev{E(t)}},
\end{equation}
with $|\tau_0| \ll t / a^2$.
This $t$-shift $\tau_0$ can be either positive or negative.
In the continuum limit $\tau_0 a^2 \to 0$ and $\gtGF(\mu) = \gGF(\mu)$.
For $\cO(a)$-improved actions like those we use, a simple calculation shows that it is possible to choose an optimal $\tau_0$ value \topt such that the $t$-shift removes the $\cO(a^2)$ corrections of the coupling $\gtGF(\mu; a)$ defined in \eq{eq:t-shift}.
In our previous studies of both the 4- and 12-flavor SU(3) systems~\cite{Cheng:2014jba}, this \topt depended only weakly on $\gtGF(\mu)$, and simply setting it to a constant value sufficed to remove most observable lattice artifacts throughout the ranges of couplings we explored in each case.

Since the gradient flow is evaluated through numerical integration, replacing $\gc \to \gtc$ by shifting $t \to t + \tau_0$ does not require any additional computation.
The $t$-shift also does not interfere with the perturbative correction in \eq{eq:pert_g2}, and in the following we will combine both improvements, searching for the optimal \topt after applying the tree-level perturbative corrections.
Using the resulting \gtc gradient flow running coupling, we will investigate the 8-flavor discrete \be function corresponding to scale change $s$,
\begin{equation}
  \label{eq:beta}
  \be_s(\gtc; L) = \frac{\gtc(sL; a) - \gtc(L; a)}{\log(s^2)}.
\end{equation}
This quantity is sometimes called the step-scaling function $\si_s(u, L)$ with $u \equiv \gtc(L; a)$, and we will use these terms interchangeably.
Our final results for the continuum discrete \be function $\be_s(\gtc) = \lim_{(a / L) \to 0} \be_s(\gtc, L)$ are then obtained by extrapolating $(a / L) \to 0$.
We emphasize that different values of $\tau_0$ should all produce the same $\be_s(\gtc)$ in the continuum limit~\cite{Cheng:2014jba}.
In \secref{sec:results} we will see that this is not actually the case for one of the lattice actions we consider.
With two nHYP smearing steps the continuum extrapolations with different $t$-shifts disagree by statistically significant amounts.
We will account for these discrepancies as one source of systematic uncertainty.

\section{\label{sec:setup}Numerical setup and lattice ensembles} 
We carry out numerical calculations using nHYP-smeared staggered fermions with smearing parameters $\al = (0.5, 0.5, 0.4)$ and either one or two smearing steps.
The gauge action includes fundamental and adjoint plaquette terms with couplings related by $\be_A / \be_F = -0.25$.
We keep the fermions exactly massless, which freezes the topological charge at $Q = 0$.
We impose anti-periodic BCs for the fermions in all four directions, but the gauge fields are periodic.

\begin{figure}[btp]
  \includegraphics[width=0.45\textwidth]{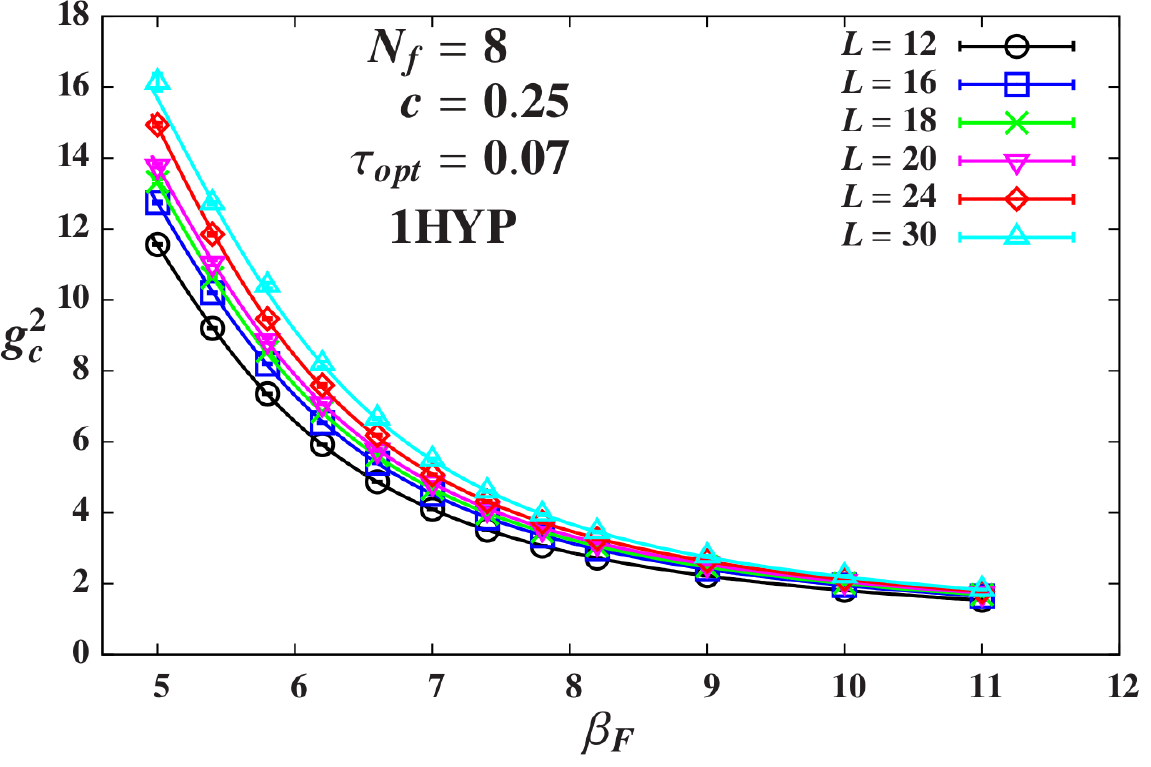}\hfill
  \includegraphics[width=0.45\textwidth]{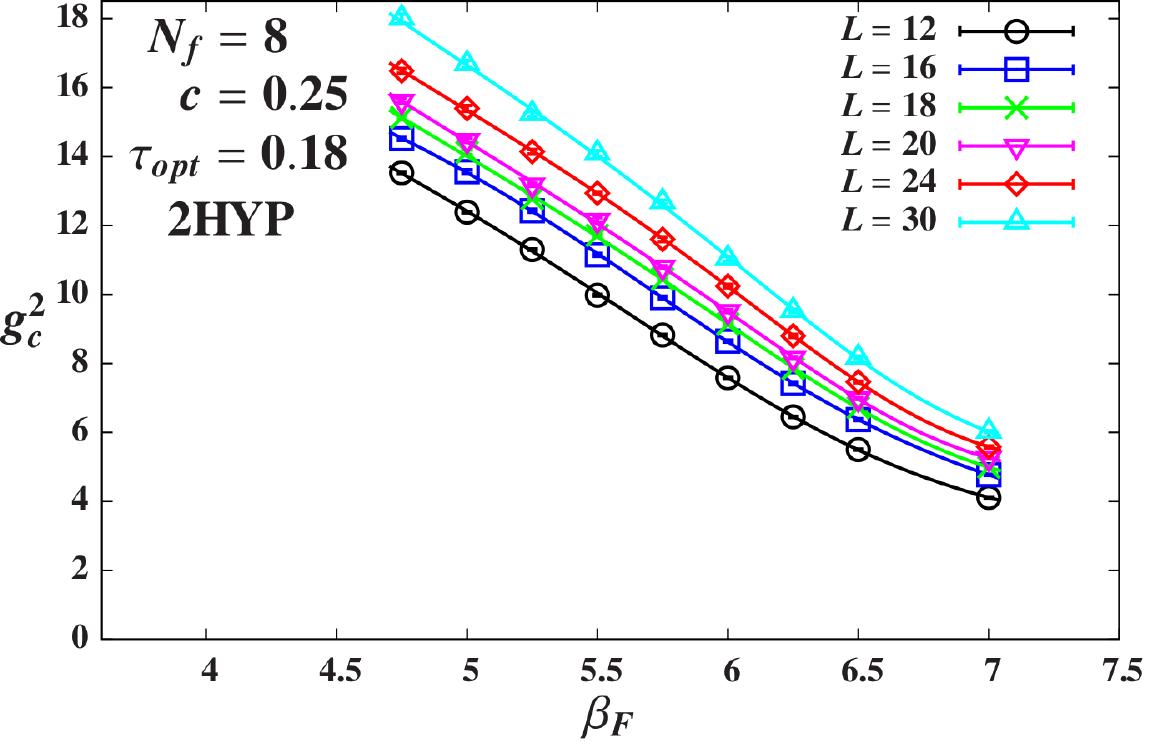}
  \caption{\label{fig:gcSq}Gradient flow renormalized coupling $\gtc(L)$ vs.\ $\be_F$ for $c = 0.25$, with optimal $\topt = 0.07$ for one nHYP smearing step (left) and $\topt = 0.18$ for two nHYP smearing steps (right).  The lines are interpolations to the (2, 2) rational function form of \protect\eq{eq:pade}.  The left edge of each plot indicates the boundary of the \Sb phase, $\be_F^{(c)} \approx 4.6$ with one smearing step and $\be_F^{(c)} \approx 3.6$ with two.}
\end{figure}

Previous studies of this lattice action with one nHYP smearing step observed an ``$\Sb$'' lattice phase in which the single-site shift symmetry ($S^4$) of the staggered action is spontaneously broken~\cite{Cheng:2011ic, Schaich:2012fr, Hasenfratz:2013uha}.
In the massless limit, a first-order transition into the \Sb phase occurs at $\be_F^{(c)} \approx 4.6$.
The twice-smeared action also has an \Sb phase that is separated from the weak-coupling phase around $\be_F^{(c)} \approx 3.6$.
In this work we consider only weaker couplings safely distant from the \Sb lattice phase.
Although the bare couplings $\be_F$ for these two different lattice actions are not directly comparable, we find that two smearing steps do allow us to access stronger renormalized couplings before encountering the \Sb phase (cf.\ \fig{fig:gcSq}).
This is consistent with our expectations; the possibility of probing stronger couplings was our main motivation for investigating the twice-smeared action in addition to the once-smeared case.
Another benefit of considering two lattice actions is that we obtain two independent sets of results.
In the continuum limit both analyses should predict the same discrete \be function, so by comparing our final results from the two different actions we can check for systematic errors.

Using each action, we generate ensembles of gauge configurations with six different $L^4$ lattice volumes with $L = 12$, 16, 18, 20, 24 and 30.
These volumes provide three pairs with scale change $s = 3 / 2$: $L = 12 \to 18$, $16 \to 24$ and $20 \to 30$.
For each $L$, with one nHYP smearing step we study twelve couplings in the range $5 \leq \be_F \leq 11$; with two smearing steps we study nine couplings in the range $4.75 \leq \be_F \leq 7$.
The resulting ensembles (72 with one smearing step and 54 with two) are summarized in Tables~\ref{tab:once-smeared} and \ref{tab:twice-smeared} in the appendix, respectively.
In \fig{fig:gcSq} we show the gradient flow renormalized coupling $\gtc(L)$ measured on each ensemble for $c = 0.25$.
These data use the optimal $t$-shift values \topt determined in the next section, and also include the tree-level perturbative correction factor $C(L, c)$ in \eq{eq:pert_g2}.
The perturbative corrections are fairly mild for the plaquette gauge action we use for both lattice generation and gradient flow, and the clover operator we use to define the energy density.
They range from $C(12, 0.25) \approx 0.947$ to $C(30, 0.25) \approx 0.981$, with similar values for the larger $c = 0.3$ and 0.35 we will also consider in \secref{sec:results}.

We can run hybrid Monte Carlo (HMC) configuration generation in the $am = 0$ chiral limit with reasonable step sizes even at the strongest bare couplings we investigate.
These $\be_F$ correspond to fairly large renormalized couplings.
For example, with $L = 30$ and one nHYP smearing step, at $\be_F = 5.0$ we have $\gtc(L) = 16.13(27)$ for $c = 0.25$, which increases to $\gc(L) = 16.43(25)$ with $\tau_0 = 0$ in \eq{eq:t-shift}.
Similarly, $\gtc(L) = 18.01(12)$ and $\gc(L) = 18.87(13)$ at $\be_F = 4.75$ with two smearing steps.
In both cases we obtain good HMC acceptance and reversibility with unit-length molecular dynamics trajectories and step sizes $\de\tau = 0.125$ at the outer level of our standard multi-timescale Omelyan integrator.
While the performance of the HMC algorithm is not a robust means to identify the phase structure of the system, this behavior indicates that none of our ensembles exhibit chiral symmetry breaking.
This conclusion is supported by our observation of a gap in the Dirac operator eigenvalue spectrum (with one nHYP smearing step) even on larger volumes at stronger $\be_F$~\cite{Cheng:2013eu, Hasenfratz:2013uha}.

Although we do reach stronger renormalized couplings with two smearing steps, the gain is fairly modest, only $\sim$15\% with $\tau_0 = 0$ and less after $t$-shift improvement.
As shown in \fig{fig:gcSq}, however, a good deal of freedom remains to extend the twice-smeared runs to stronger couplings before encountering the \Sb phase, which is located at the left edge of each plot.
The computational cost of such runs prevents us from including them in the present work.
As tabulated in Tables~\ref{tab:once-smeared} and \ref{tab:twice-smeared}, the gauge fields generated in the twice-smeared runs are already quite rough, with average plaquettes approaching $1 / 3$.
In addition, as we will show below our twice-smeared results already exhibit cutoff effects significantly larger than those we observe with one smearing step, suggesting that pushing this action to stronger couplings may not be worth the computational expense.

\section{\label{sec:results}Step-scaling analyses and results} 
Following the standard procedure for lattice step-scaling analyses, we will first fit our input data to some interpolating function to determine the finite-volume discrete \be functions $\be_s(\gtc, L)$ with fixed $L$ (\eq{eq:beta}), and then extrapolate these to the $(a / L)^2 \to 0$ continuum limit.
Because we consider the same input bare couplings $\be_F$ (giving the same lattice spacings $a$) on every lattice volume, we can either interpolate the renormalized couplings $\gtc(L)$ as functions of $\be_F$, or at each input $\be_F$ we can compute $\be_s(\gtc, L)$ directly from \eq{eq:beta} and interpolate these as functions of $\gtc(L)$.
We will carry out analyses using both approaches, and interpret any disagreement between them as a systematic error.
A similar procedure was used by \refcite{Fodor:2012td}.
In this work, we find that our results from the two approaches always agree within statistical uncertainties.

\begin{figure}[btp]
  \includegraphics[width=0.45\textwidth]{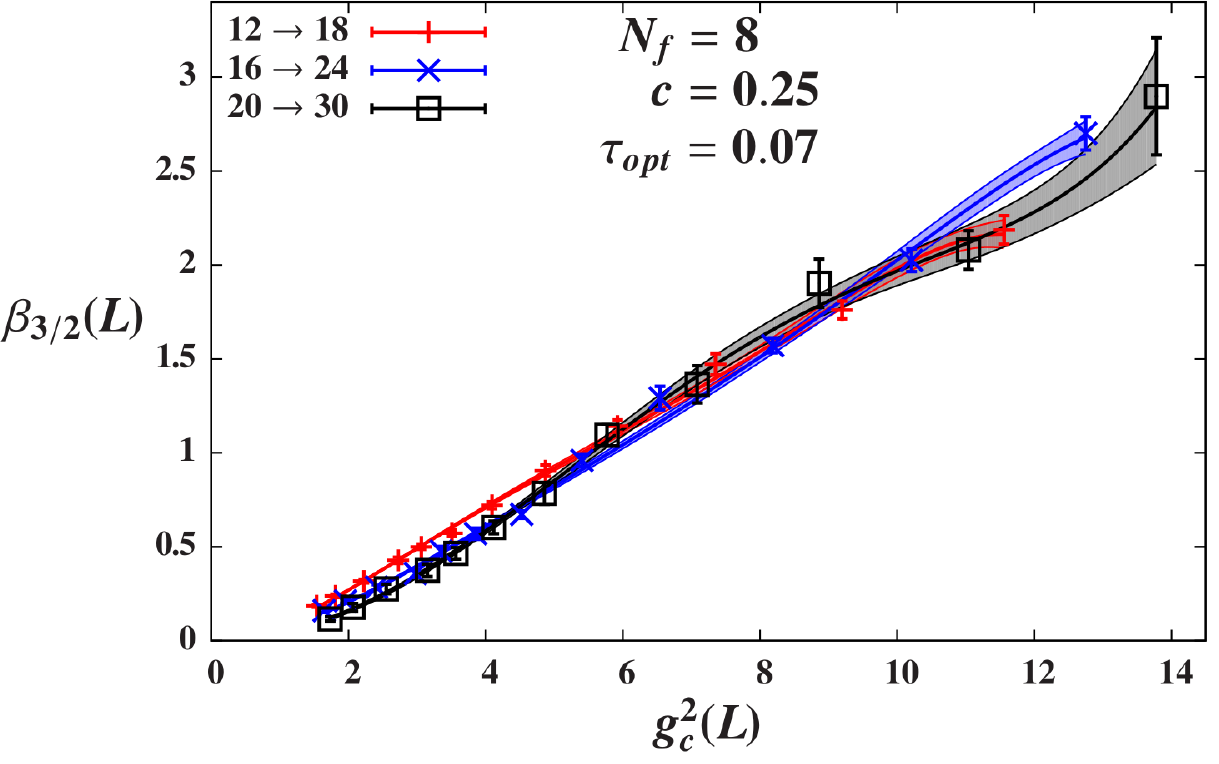}\hfill
  \includegraphics[width=0.45\textwidth]{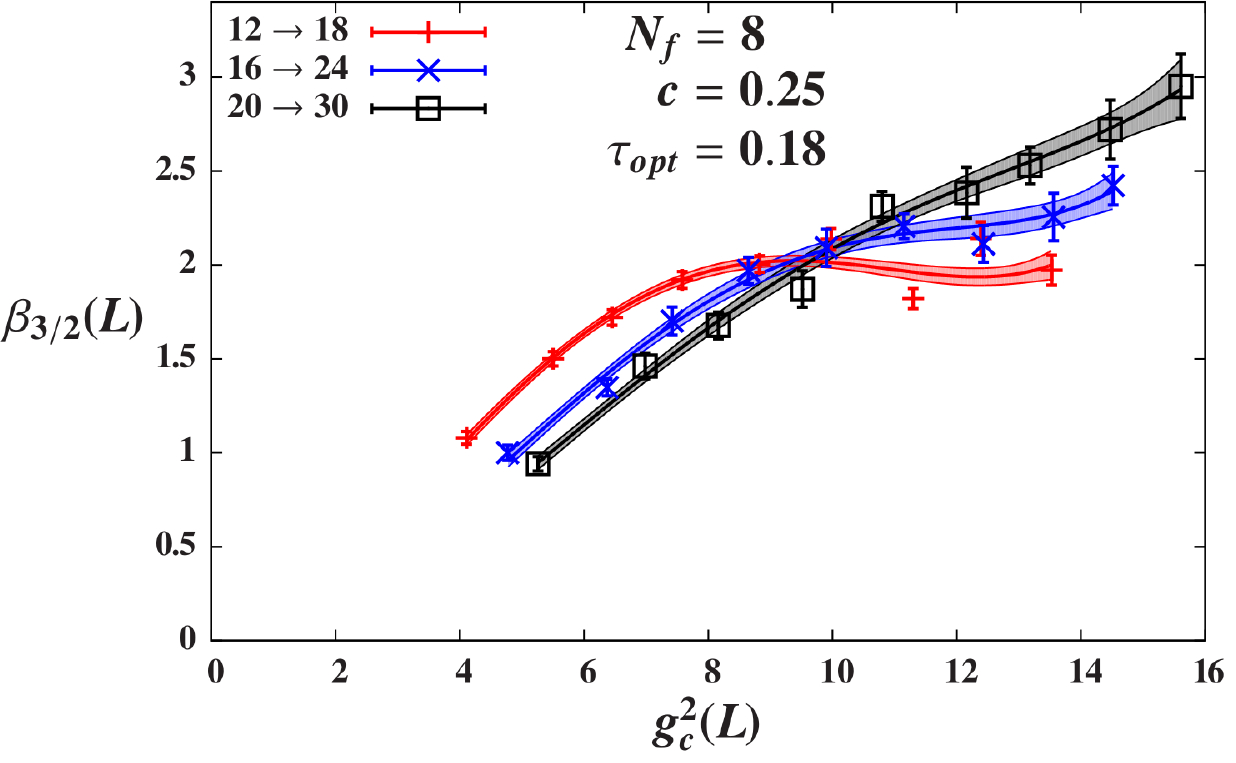}
  \caption{\label{fig:betaL}Finite-volume discrete \be function $\be_s(L) = \left[\gtc(sL) - \gtc(L)\right] / \log(s^2)$ vs.\ $\gtc(L)$ for $c = 0.25$ and $s = 3 / 2$, with optimal $\topt = 0.07$ for one nHYP smearing step (left) and $\topt = 0.18$ for two nHYP smearing steps (right).  The lines and error bands are interpolations using the four-term polynomial form of \protect\eq{eq:poly}.  The agreement between the three different $L \to sL$ curves in the left panel indicates that $\topt = 0.07$ removes most cutoff effects for the once-smeared action, in the investigated coupling range.  In the twice-smeared case on the right, significant cutoff effects remain for any constant $\topt$, though the continuum extrapolations remain linear in $(a / L)^2$ as desired.}
\end{figure}

Fitting the renormalized coupling $\gtc(L)$ on each lattice volume to some interpolating function in the bare coupling $g_0^2 \equiv 12 / \be_F$ is the more traditional approach.
While the choice of interpolating function is essentially arbitrary, typically some functional form motivated by lattice perturbation theory is used.
For example, refs.~\cite{Appelquist:2009ty, Fodor:2012td} both fit $\frac{1}{g^2} - \frac{1}{g_0^2}$ to polynomials in $g_0^2$.
Inspired by \refcite{Karavirta:2011zg}, we instead consider rational function interpolations.
Specifically, the interpolating curves shown in \fig{fig:gcSq} use the ``(2, 2)'' rational function
\begin{equation}
  \label{eq:pade}
  \gtc(L) = \left(\frac{12}{\be_F}\right) \frac{1 + c_0 \be_F + c_1 \be_F^2}{c_2 + c_3 \be_F + c_4 \be_F^2},
\end{equation}
which reduces to the expected $\gtc \propto g_0^2$ at weak coupling.
Most of the fits shown are of good quality, with $0.2 \lsim \chidof \lsim 1.6$, corresponding to confidence levels $0.94 \gsim CL \gsim 0.16$.
The main outlier is the twice-smeared $L = 12$ interpolation, which has $\chidof \approx 3.8$ and $CL \approx 0.004$.
While the quality of fits can be improved by adjusting the number of terms in the rational function, the final results are unchanged within statistical uncertainties.

When we interpolate the finite-volume $\be_s(L)$ from \eq{eq:beta} as functions of $\gtc(L)$, it is reasonable to use the same sort of polynomial interpolating function that perturbation theory predicts for the continuum \be function,
\begin{equation}
  \label{eq:poly}
  \be_s(\gtc; L) = \frac{\gtc(sL) - \gtc(L)}{\log(s^2)} = \widetilde g_c^4(L) \sum_{i = 0}^N b_i\ \widetilde g_c^{2i}(L).
\end{equation}
In \fig{fig:betaL} we show $\be_s(L)$ data along with interpolations to four-term polynomials ($N = 3$ in \eq{eq:poly}), using the same $c = 0.25$ and \topt as in \fig{fig:gcSq}.
The quality of these fits is comparable to that of the rational function fits discussed above, with typical $0.4 \lsim \chidof \lsim 1.4$ and $0.93 \gsim CL \gsim 0.18$.
Again, the twice-smeared $L = 12$ data are not well behaved, with $\chidof \approx 3.4$, $CL \approx 0.004$ and deviations clearly visible in \fig{fig:betaL}.
Since the polynomial functional form of \eq{eq:poly} is better motivated than the rational function in \eq{eq:pade}, and additionally involves only renormalized and not bare quantities, we will take our final results from this analysis, using the rational function analysis only to determine (vanishing) contributions to the systematic errors.

\begin{figure}[btp]
  \includegraphics[width=0.3\textwidth]{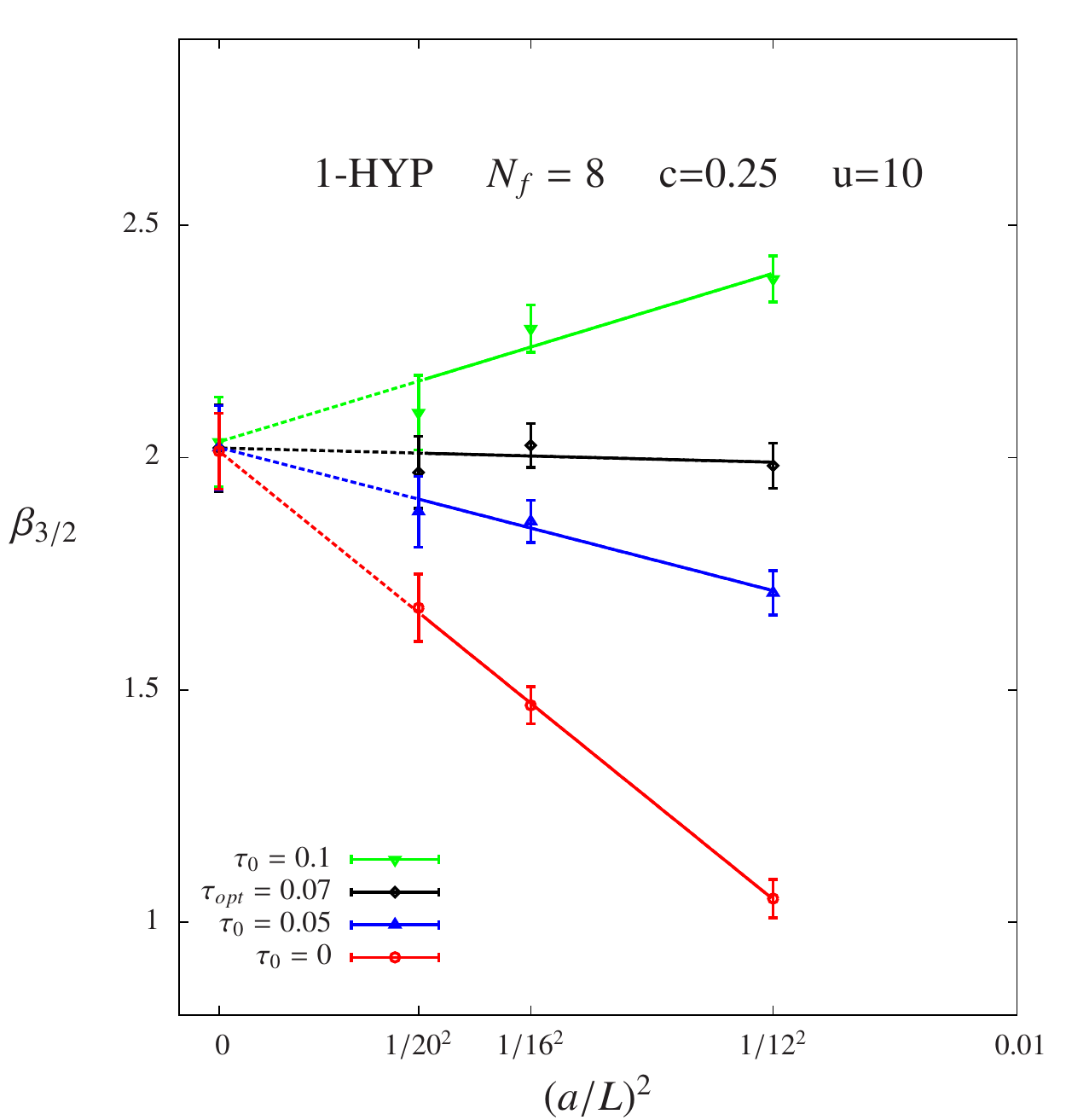}\hfill
  \includegraphics[width=0.3\textwidth]{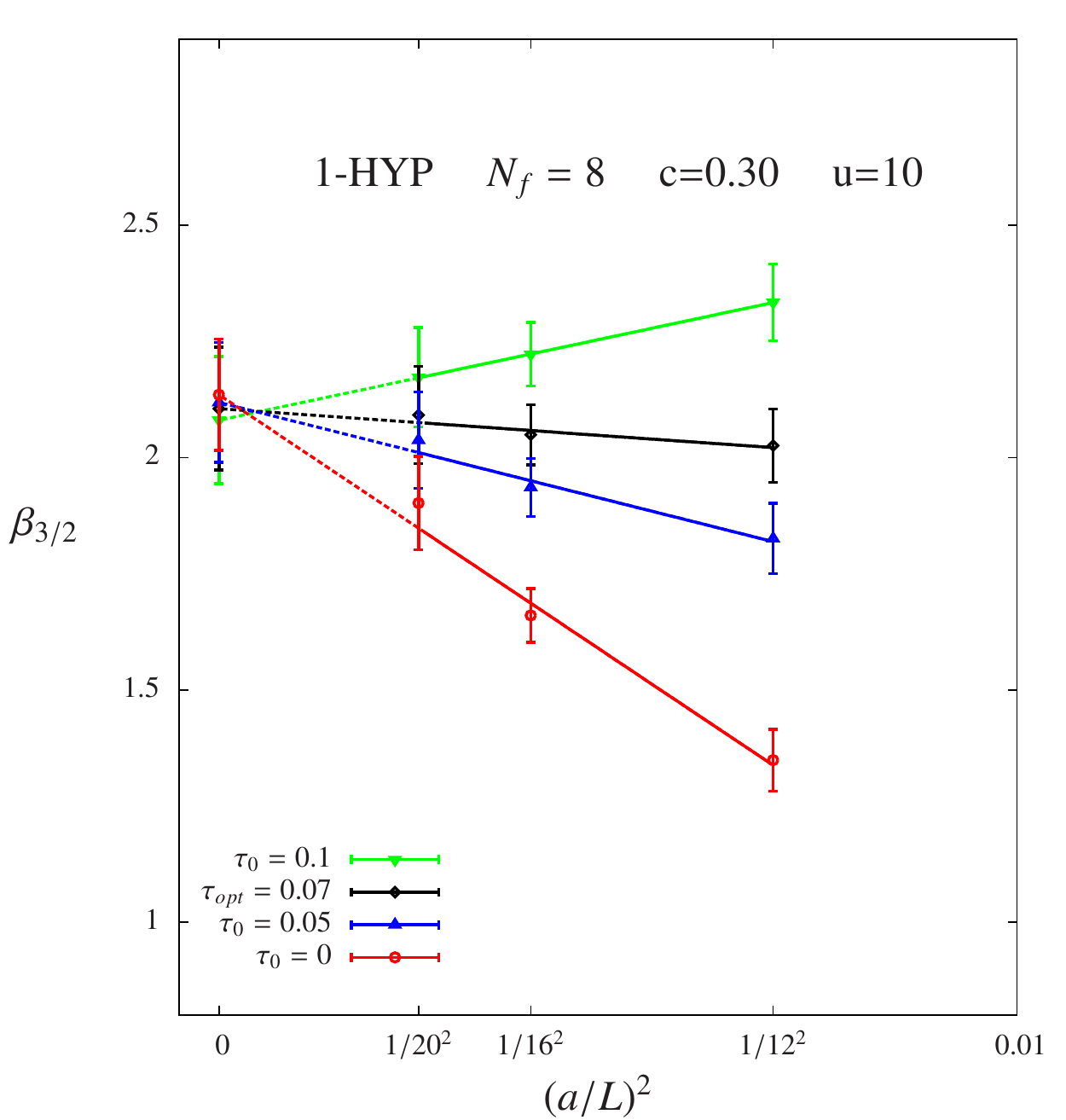}\hfill
  \includegraphics[width=0.3\textwidth]{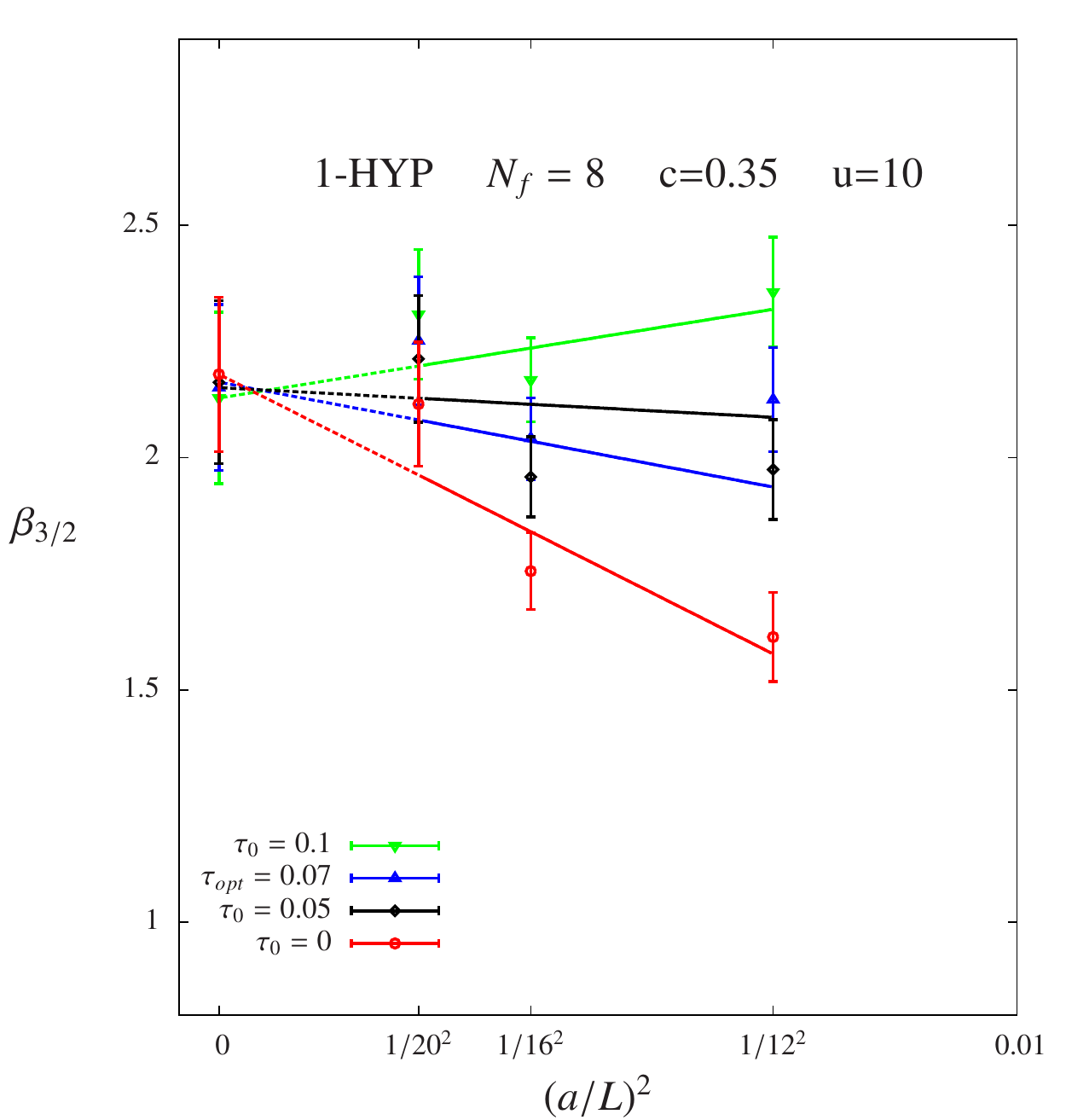}
  \caption{\label{fig:extrap_1HYP}Linear $(a / L)^2 \to 0$ continuum extrapolations of the once-smeared $s = 3 / 2$ discrete \be function for $u = 10$ and $c = 0.25$ (left), 0.3 (center) and 0.35 (right).  In each plot we compare several values of $\tau_0 = 0$, 0.05 and 0.1, in addition to the optimal $\topt = 0.07$.} \vspace{12 pt}
  \includegraphics[width=0.3\textwidth]{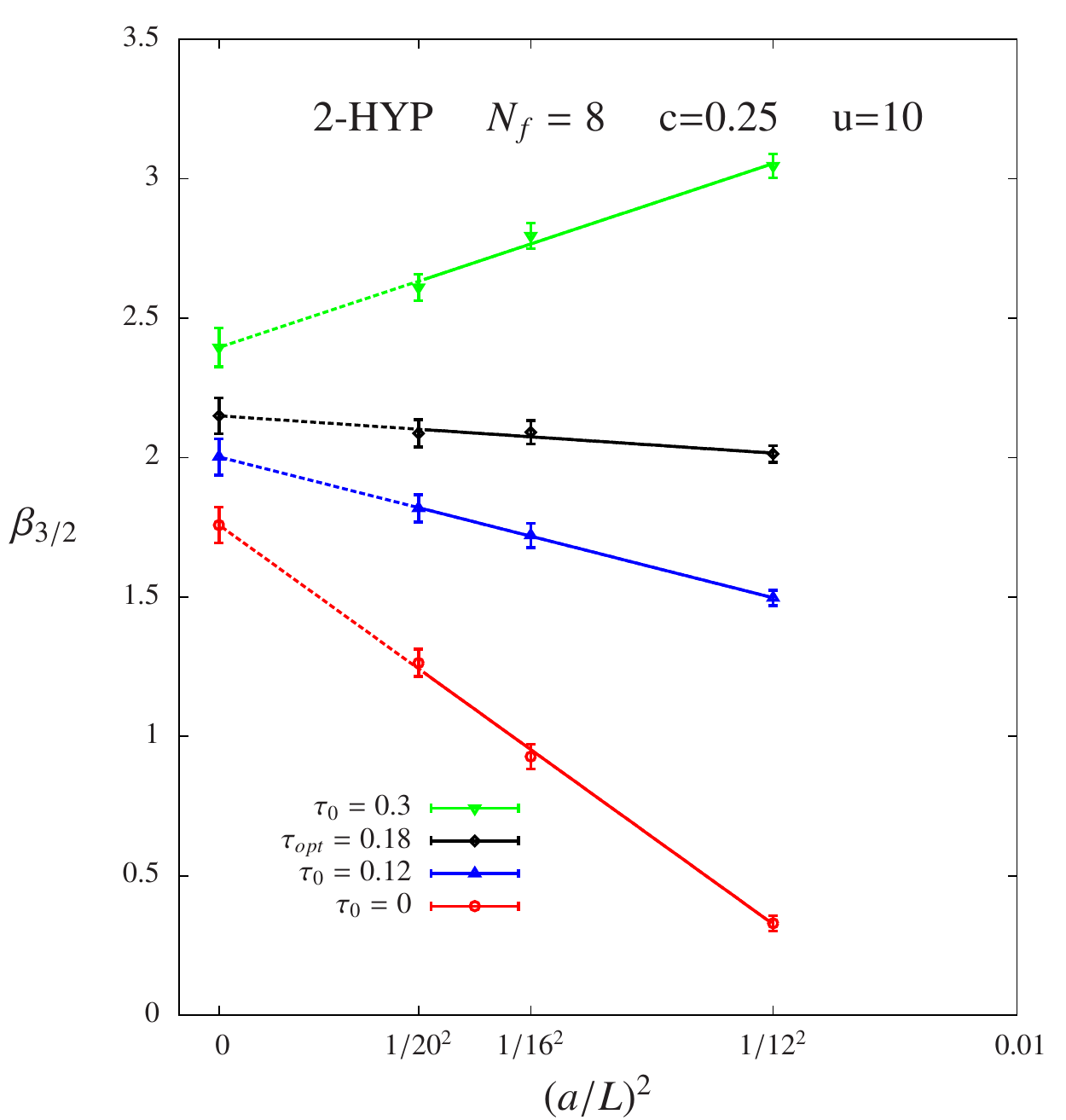}\hfill
  \includegraphics[width=0.3\textwidth]{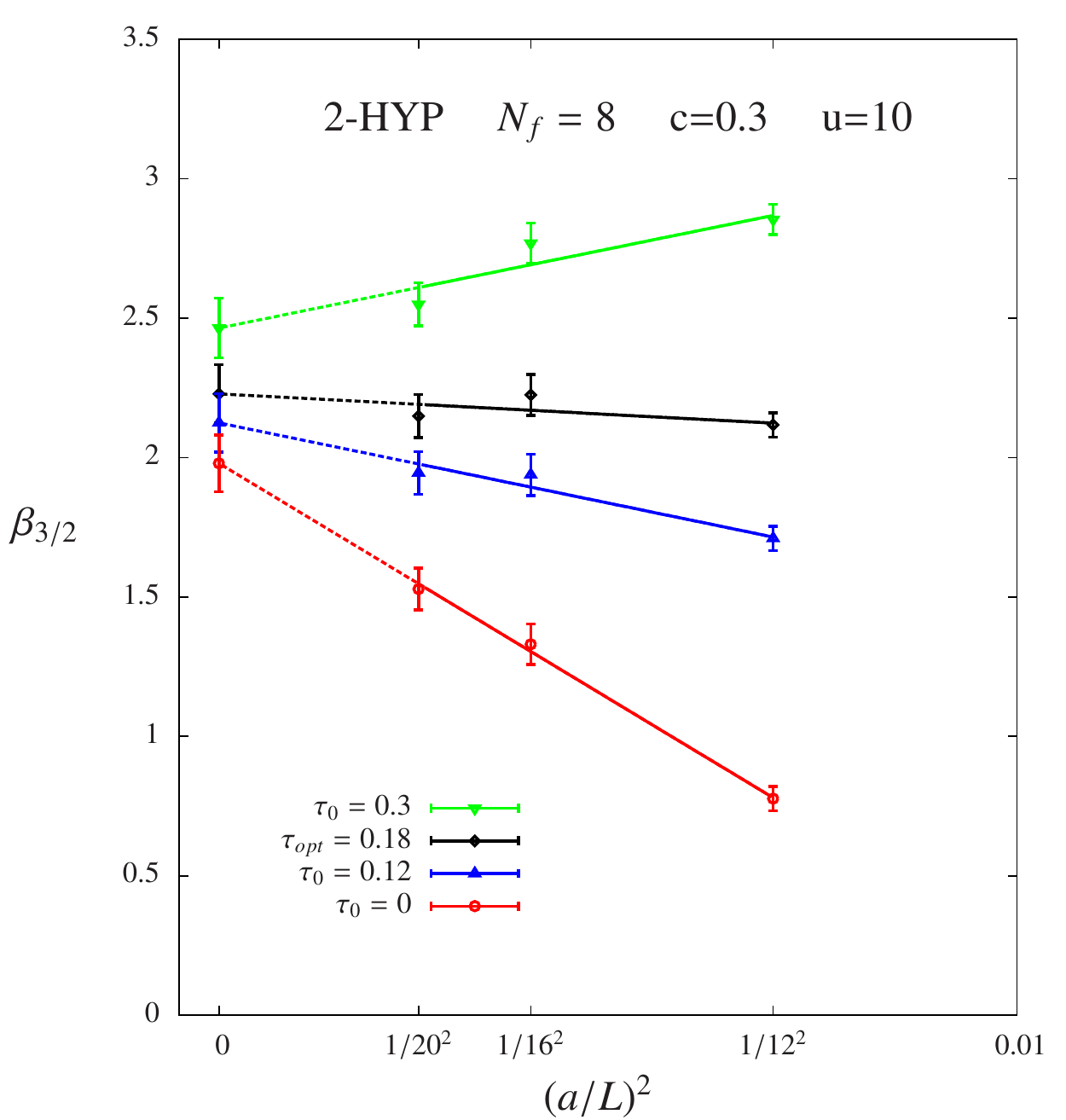}\hfill
  \includegraphics[width=0.3\textwidth]{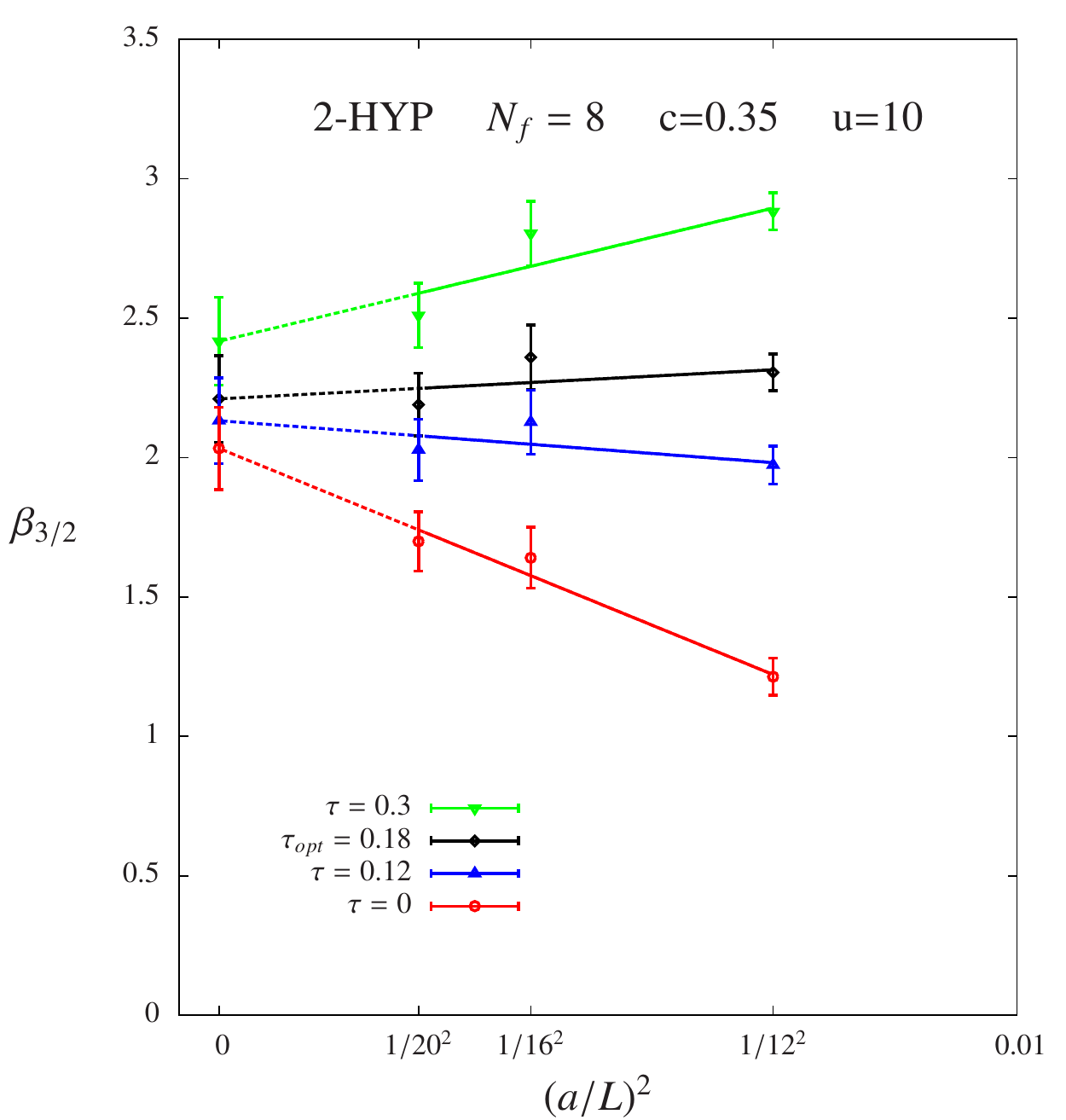}
  \caption{\label{fig:extrap_2HYP}Linear $(a / L)^2 \to 0$ continuum extrapolations of the twice-smeared $s = 3 / 2$ discrete \be function for $u = 10$ and $c = 0.25$ (left), 0.3 (center) and 0.35 (right).  In each plot we compare several values of $\tau_0 = 0$, 0.12 and 0.3, in addition to the optimal $\topt = 0.18$.}
\end{figure}

Now we turn to the continuum extrapolation and the related issue of optimizing the $t$-shift $\tau_0$.
Since staggered fermions are $\cO(a)$ improved, we extrapolate linearly in $(a / L)^2 \to 0$ to the continuum limit.
Several representative extrapolations with fixed $u = 10$ are shown in figures~\ref{fig:extrap_1HYP} and \ref{fig:extrap_2HYP}, for the once- and twice-smeared actions, respectively.
Each figure contains separate panels for $c = 0.25$, 0.3 and 0.35, each of which considers several values of $\tau_0$, including $\tau_0 = 0$.
These continuum extrapolations all follow the expected linear dependence on $(a / L)^2$, with average confidence levels of 0.49 with one nHYP smearing step and 0.46 with two.
As illustrated by figure~4 of \refcite{Fritzsch:2013je}, $(a / L)^2$ scaling should become more reliable as $c$ increases within the range we consider. 
At the same time, however, larger values of $c$ produce larger statistical uncertainties.
Since the left panels of both figures~\ref{fig:extrap_1HYP} and \ref{fig:extrap_2HYP} show linear $(a / L)^2$ dependence with $c = 0.25$, we have no reason to doubt that the stronger fluctuations observed with $c = 0.35$ in the right panels of these figures are due to statistics.

In every case the unshifted $\tau_0 = 0$ results show significant dependence on $(a / L)^2$, despite the tree-level perturbative correction discussed in \secref{sec:gradflow}.
We wish to optimize $\tau_0$ by finding the value \topt for which these cutoff effects are minimized.
As discussed in \secref{sec:gradflow}, we will use constant \topt for all $\gtc$, which will only reduce and not completely remove $\cO(a^2)$ effects.
With one nHYP smearing step, our choice $\topt = 0.07$ is satisfactory for all couplings we consider.
\Fig{fig:extrap_1HYP} shows the resulting removal of cutoff effects for one particular $u = 10$, while the left panel of \fig{fig:betaL} considers the full range of \gtc with $c = 0.25$.
The three curves in \fig{fig:betaL} are the finite-volume discrete \be functions that we extrapolate to the continuum, and with $\topt = 0.07$ they nearly overlap for all couplings.

The twice-smeared action is quite different.
In this case we choose $\topt = 0.18$, more than 2.5 times larger than the once-smeared $\topt = 0.07$, which already indicates more severe cutoff effects.
While this $\topt = 0.18$ produces the desired nearly constant continuum extrapolations for $u = 10$ in \fig{fig:extrap_2HYP}, from the right panel of \fig{fig:betaL} we can see that cutoff effects remain for both smaller and larger couplings.
Specifically, a smaller $t$-shift $\tau_0 \approx 0.12$ produces better improvement for $u \lsim 8$, while a larger $\tau_0 \approx 0.24$ is more effective for $u \gsim 12$.
Our choice of constant $\topt = 0.18$ is a compromise that ``over-improves'' at small \gtc and ``under-improves'' at large $\gtc$.
While it is possible to use a $u$-dependent $\topt$, we prefer to keep this improvement as simple as possible, to avoid the risk of losing predictivity by introducing too many optimization parameters~\cite{Cheng:2014jba}.
Although we end up with non-trivial continuum extrapolations as indicated by the right panel of \fig{fig:betaL}, they remain reliably linear in $(a / L)^2$. 

The fact that we observe larger lattice artifacts as a result of performing a second nHYP smearing step may seem surprising at first glance, since smearing is typically expected to reduce such artifacts.
This behavior can be understood by recalling that we are attempting to probe the 8-flavor system at its strongest accessible couplings.
In this regime, in order to end up with large renormalized couplings after two smearing steps we must work with extraordinarily rough unsmeared gauge links.
These rough unsmeared links appear in the gauge action and may be blamed for the large lattice artifacts, the large value of $\topt = 0.18$, and the remaining $\cO(a^2)$ effects visible in \fig{fig:betaL} even after $t$-shift improvement.

Of greater concern than these remaining $\cO(a^2)$ effects is the fact that the large lattice artifacts of the twice-smeared action can affect even the continuum-extrapolated discrete \be function results when using lattice volumes $12 \leq L / a \leq 30$.
Reliable continuum extrapolations should behave as shown for the once-smeared action in \fig{fig:extrap_1HYP}, where different values of $\tau_0$ predict the same $(a / L)^2 \to 0$ limit $\be_s(\gtc)$ well within statistical uncertainties.
In this case the $t$-shift improvement simply stabilizes the extrapolations by removing cutoff effects, without changing the continuum results.
The contrast with \fig{fig:extrap_2HYP} for two nHYP smearing steps is dramatic, especially for smaller $c$.
In this case different $t$-shifts produce continuum-extrapolated $\be_s(\gtc)$ that disagree by statistically significant amounts, indicating considerable systematical errors in continuum extrapolations using the available lattice volumes without the $t$-shift improvement.
There are also additional sources of systematic uncertainties, which we will now discuss.

\begin{figure}[btp]
  \includegraphics[width=0.45\textwidth]{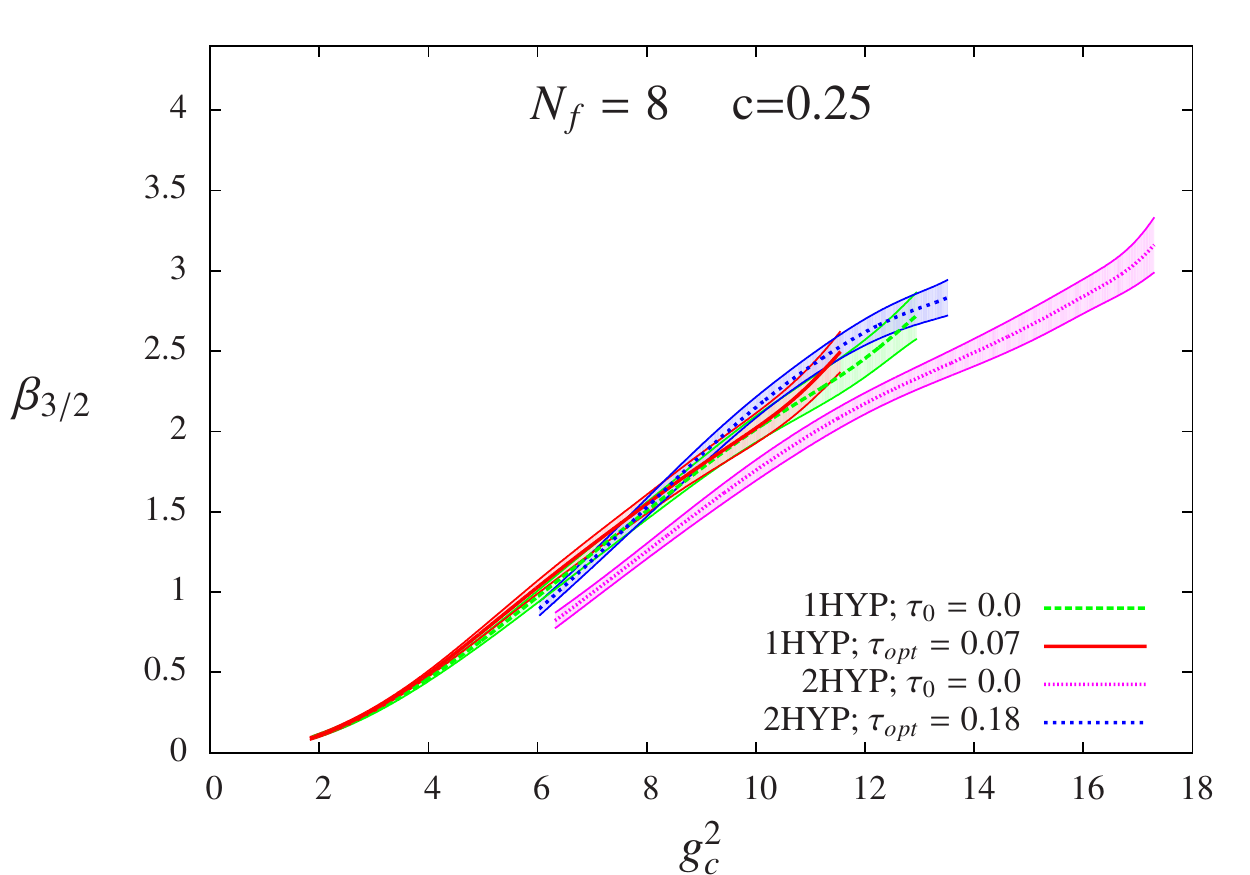}\hfill
  \includegraphics[width=0.45\textwidth]{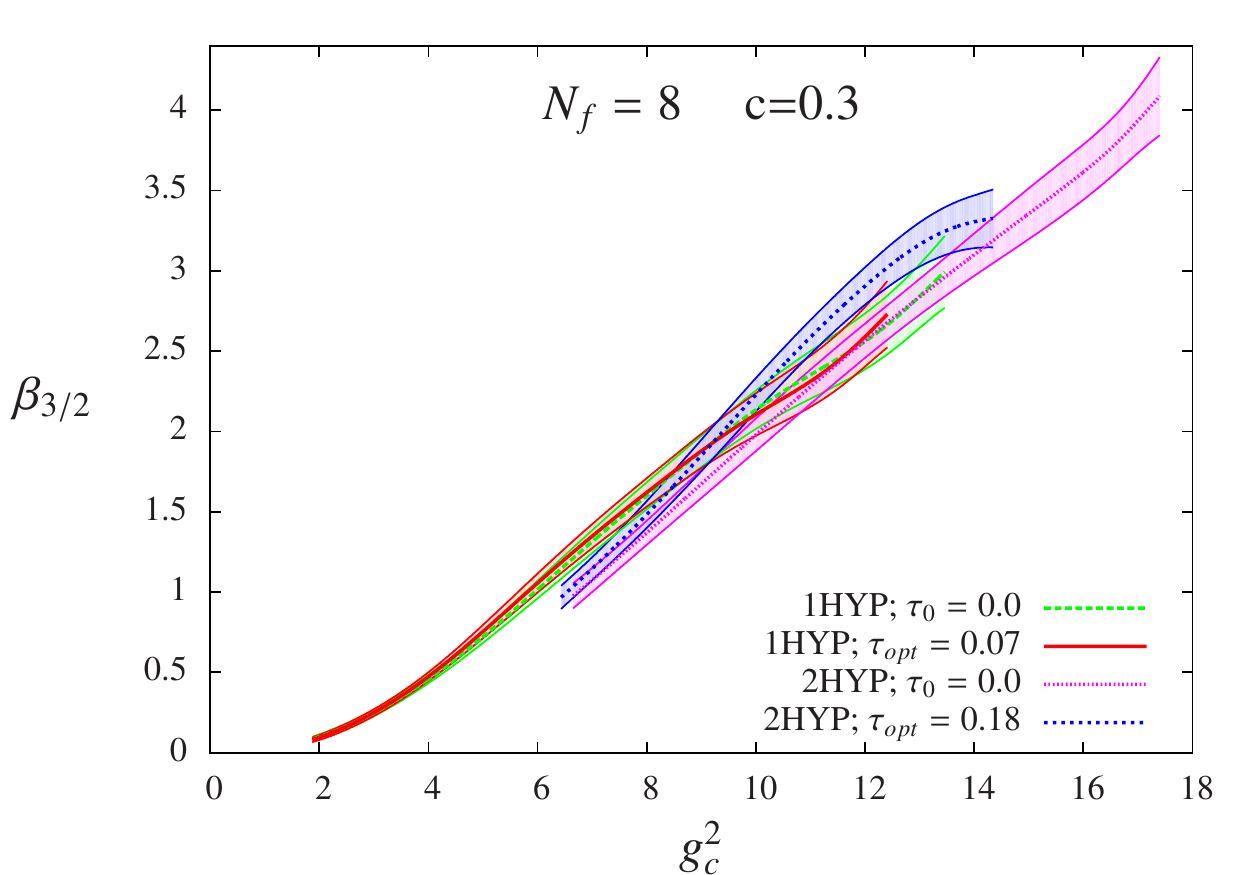}
  \caption{\label{fig:beta_opt}Continuum-extrapolated discrete \be function for scale change $s = 3 / 2$ with $c = 0.25$ (left) and 0.3 (right).  In each plot we include once- and twice-smeared results with $\tau_0 = 0$ as well as with the optimal $\topt = 0.07$ and 0.18, respectively.  While $\tau_0$ optimization can change the twice-smeared continuum limit, removing a source of systematic error, the once-smeared results always agree within uncertainties.}
\end{figure}

We account for three potential sources of systematic errors:
\begin{description}
  \setlength{\itemsep}{1 pt}
  \setlength{\parskip}{0 pt}
  \setlength{\parsep}{0 pt}
  \vspace{-6 pt}
  \item[Optimization:] To determine how we should account for any sensitivity to the $t$-shift improvement parameter $\tau_0$, consider \fig{fig:beta_opt}.
    Each panel in this figure compares once- and twice-smeared continuum-extrapolated results with both $\tau_0 = 0$ and the optimal $\topt$, which should all predict the same $\be_s(\gtc)$.
    While the once-smeared results always agree within uncertainties, optimizing $\tau_0$ produces a statistically significant change with two nHYP smearing steps, just as in \fig{fig:extrap_2HYP}.
    In fact, the $t$-shift brings the twice-smeared results into better agreement with the once-smeared action, {\em removing} systematic errors that would be present for an unimproved analysis with $\tau_0 = 0$.

    The only remaining systematic uncertainties from optimization therefore result from our restriction to constant $\topt$.
    As discussed above, $\topt = 0.07$ is satisfactory for all $\gtc$, so these systematic uncertainties vanish for the once-smeared action.
    With two nHYP smearing steps, however, $\tau_0 = 0.12$ (0.24) is preferred for small (large) $\gtc$.
    We conservatively define as systematic errors any discrepancies between results for either of these two $\tau_0$ compared to those for $\topt = 0.18$.
    These systematic errors tend to be quite mild, at least 3.5 times smaller than the statistical uncertainties.

  \item[Interpolation:] As discussed at the start of this section, we analyze our data both by interpolating $\gtc(L)$ as functions of $\be_F$ and by interpolating $\be_s(\gtc, L)$ as functions of $\gtc(L)$.
    We take our final results from the latter analysis.
    Any discrepancies between the two approaches we include as a systematic error.
    For the 8-flavor analyses we carry out in this work, these systematic errors always vanish.

  \item[Extrapolation:] Even after accounting for tree-level perturbative corrections and $t$-shift improvement, our continuum extrapolations are not always perfectly linear in $(a / L)^2$.
    To determine the resulting systematic effects, we repeat all analyses without including the smallest-volume $L = 12 \to 18$ data, considering only $16 \to 24$ and $20 \to 30$ points in linear $(a / L)^2 \to 0$ extrapolations.
    Any discrepancies between the two- and three-point continuum extrapolations defines our third systematic uncertainty.
    Although this source of systematic error also often vanishes, for some $u$ it can be up to four times larger than the statistical uncertainty.
\end{description}
\vspace{-6 pt}
In all three cases, we take the systematic errors to vanish when the results being compared agree within 1$\si$ statistical uncertainties.
This ensures that statistical fluctuations are not double-counted as both systematic and statistical errors.
Note that to determine the systematic uncertainties from $\tau_0$ optimization, it was important to compare multiple lattice actions.
We will return to this point in \secref{sec:conclusion}.

\begin{figure}[btp]
  \includegraphics[width=0.45\textwidth]{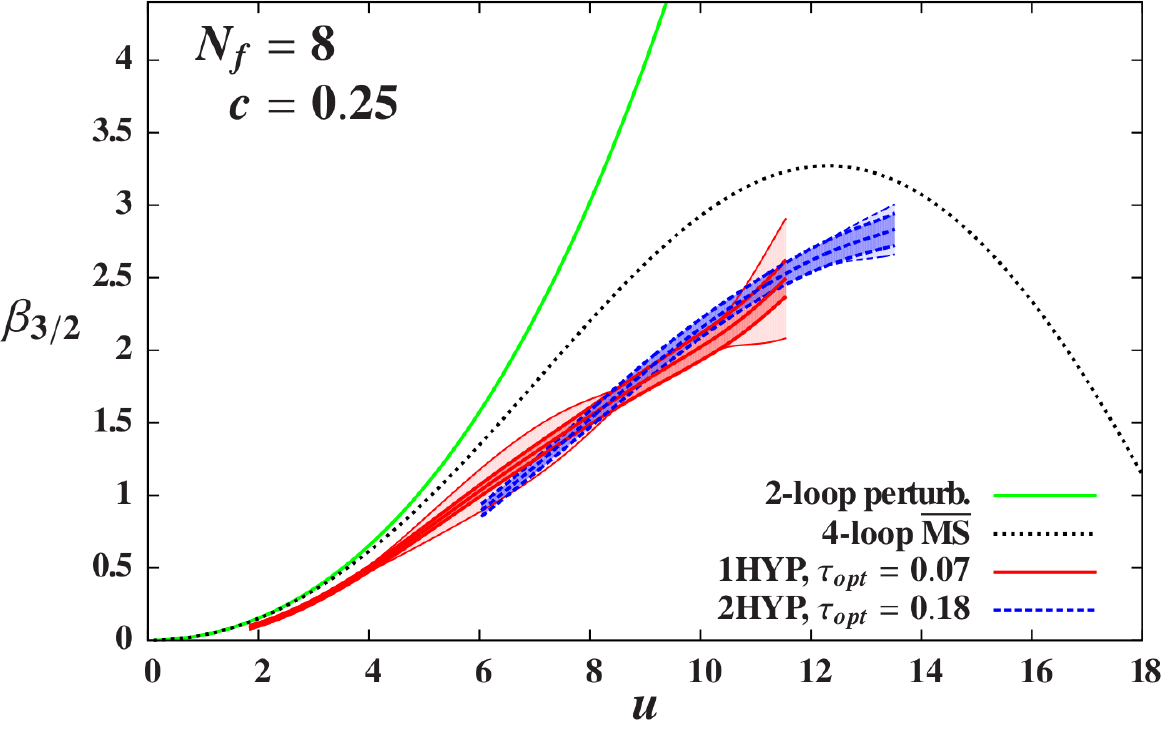}\hfill
  \includegraphics[width=0.45\textwidth]{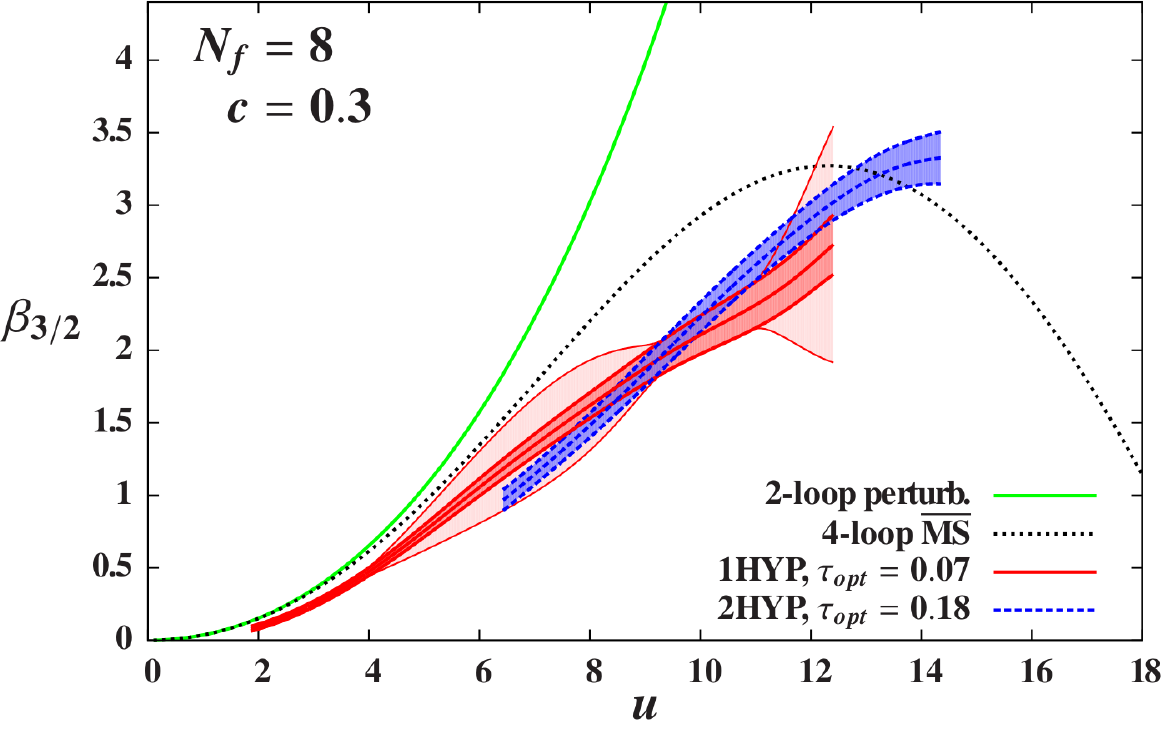}
  \caption{\label{fig:beta}Continuum-extrapolated discrete \be function for scale change $s = 3 / 2$ with $c = 0.25$ (left) and 0.3 (right).  In each plot we include once- and twice-smeared results using the optimal $\topt = 0.07$ and 0.18, respectively, as well as two-loop perturbation theory (solid line) and the four-loop perturbative prediction in the \MSbar scheme (dotted line).  The darker error bands indicate our statistical uncertainties, while the lighter error bands show the total uncertainties, with statistical and systematic errors added in quadrature.}
\end{figure}

We are now ready to present our final results for the 8-flavor system.
\Fig{fig:beta} shows the continuum-extrapolated $s = 3 / 2$ discrete \be function for two different renormalization schemes, $c = 0.25$ and 0.3.
In both panels we include our nonperturbative results for the once- and twice-smeared actions.
The darker error bands show the statistical uncertainties, while the lighter error bands indicate the total uncertainties, with statistical and systematic errors added in quadrature.

We compare our numerical results with perturbation theory, where
\begin{align}
  & L^2 \frac{dg^2}{dL^2} = \frac{g^4}{16\pi^2} \sum_{i = 0} b_i \left(\frac{g^2}{16\pi^2}\right)^i \\
  b_0 & = \frac{11}{3}C_2(G) - \frac{4}{3}N_f T(R) \cr
  b_1 & = \frac{34}{3}\left[C_2(G)\right]^2 - N_f T(R) \left[\frac{20}{3}C_2(G) + 4C_2(R)\right] \nn
\end{align}
for $N_f$ fermions transforming in representation $R$ of the gauge group.
For the fundamental representation of SU(3) gauge theory,
\begin{align}
  C_2(G) & = 3 &
  T(F) & = \frac{1}{2} &
  C_2(F) & = \frac{4}{3},
\end{align}
so that $N_f = 8$ gives $b_0 = \frac{17}{3}$ and $b_1 = \frac{2}{3}$.
Higher-order coefficients $b_i$ are renormalization scheme dependent.
In the \MSbar scheme, \refcite{Ryttov:2010iz} reports numerical values $b_2 \approx -423$ and $b_3 \approx 374$ for 8-flavor SU(3) gauge theory.
Both the three- and four-loop \be functions predict an IR fixed point, but only at strong couplings $g_{\MSbar}^2 \approx 18.4$ and 19.5 where perturbation theory is not reliable.

Along with our numerical results we include the two- and four-loop perturbative predictions for the $s = 3 / 2$ discrete \be function in \fig{fig:beta}.
The once- and twice-smeared actions predict consistent continuum results, which are significantly smaller than the two-loop perturbative curve, by more than a factor of three for $\gtc = 12$.
At the weakest coupling that we probe, $\gtc \approx 2$, our results are still approaching the perturbative predictions from below.
(Although we mentioned in \secref{sec:gradflow} that the gradient flow discrete \be function is only one-loop universal, the one- and two-loop perturbative results are almost indistinguishable across the range shown in \fig{fig:beta}.)
Due to the large negative $b_2$ coefficient in the \MSbar scheme, the four-loop discrete \be function also becomes much smaller than the two-loop prediction.
Even at the strongest $\gtc = 13.5$ that we are able to reach with two nHYP smearing steps in the $c = 0.25$ scheme, our numerical results remain even smaller than four-loop perturbation theory.
In the $c = 0.3$ scheme the twice-smeared \be function becomes comparable to the maximum of the four-loop curve at the largest accessible $\gtc = 14.3$.
Of course, since the discrete \be function is scheme dependent the $c = 0.25$ and 0.3 results do not have to agree.
The perturbative four-loop \MSbar \be function not only corresponds to another different scheme, but is also of questionable validity at such strong couplings.
Our comparisons with perturbation theory are for illustration only.

\section{\label{sec:conclusion}Discussion and conclusions} 
Before we attempt to interpret our nonperturbative results for the discrete \be function in \fig{fig:beta}, let us review the motivations for and goals of this work.
We are attracted to 8-flavor SU(3) gauge theory primarily by the possibility that it may possess strongly coupled near-conformal IR dynamics, leading to desirable BSM phenomenology including a light Higgs particle~\cite{Aoki:2014oha} and large effective mass anomalous dimension across a wide range of energy scales~\cite{Cheng:2013eu}.
While a variety of existing lattice studies have not yet been able to establish chiral symmetry breaking in the $am \to 0$ limit for $N_f = 8$, their results are all consistent with chirally broken dynamics~\cite{Appelquist:2007hu, Appelquist:2009ty, Deuzeman:2008sc, Miura:2012zqa, Jin:2010vm, Schaich:2012fr, Hasenfratz:2013uha, Fodor:2009wk, Aoki:2013xza, Aoki:2014oha, Schaich:2013eba, Appelquist:2014zsa, Cheng:2013eu}.
To address this situation, we have carried out a new step-scaling study of the discrete \be function, exploiting two different improved lattice actions and the recently introduced gradient flow running coupling that enabled us to investigate significantly stronger couplings than were previously accessible.

Our results in \fig{fig:beta} indicate a coupling that runs much more slowly than predicted by two-loop perturbation theory, even more slowly than the four-loop \MSbar prediction, which possesses a strongly coupled IR fixed point.
Despite considering a second lattice action with two nHYP smearing steps, in addition to our usual once-smeared action, we could not reach strong enough couplings either to see a similar IRFP in our numerical results, or to obtain a clear deviation from the IR-conformal four-loop result.\footnote{Recent lattice studies of the 12-flavor system have reported surprisingly close agreement with the four-loop \MSbar scheme~\cite{Ryttov:2010iz}, both for the scheme-dependent location of this system's IR fixed point~\cite{Cheng:2014jba}, and for the scheme-independent mass anomalous dimension at the IRFP~\cite{Cheng:2013xha, Lombardo:2014pda}.} 
We see no sign of spontaneous chiral symmetry breaking for running couplings as large as $\gc \approx 19$ with $\tau_0 = 0$ on $30^4$ lattice volumes.
In part, our ability to access stronger renormalized couplings is limited by the more severe lattice artifacts in our twice-smeared results.
In this case, significant $t$-shift improvement is required to obtain agreement with the once-smeared results, illustrating the importance of cross-checking continuum predictions by comparing different lattice actions.
As shown by \fig{fig:beta_opt}, the large $\topt = 0.18$ needed with two smearing steps significantly reduces the range of \gtc that we can reach on lattice volumes from $12^4$ to $30^4$.
In addition, any constant \topt necessarily leaves non-negligible cutoff effects in our twice-smeared results, in contrast to the better behavior we observe with one nHYP smearing step.
This is illustrated in \fig{fig:betaL}.
While we could address this issue by allowing \topt to depend on the bare or renormalized coupling, we prefer to keep the $t$-shift improvement as simple as possible, and to interpret the more limited improvement of the twice-smeared system as a warning that this lattice action suffers from significant artifacts.
Even though we can still push twice-smeared computations to stronger bare couplings before encountering the \Sb lattice phase (\fig{fig:gcSq}), the severe cutoff effects we observe suggest that doing so may not be worth the computational expense.

Future investigations of the 8-flavor system will benefit from several studies currently being carried out with the once-smeared action we considered in this work.
As discussed in \secref{sec:intro}, the Lattice Strong Dynamics Collaboration is studying the finite-temperature phase diagram with $N_t = 24$, which still seems to be too small to establish chiral symmetry breaking in the massless limit~\cite{LSDfiniteT}.
At the same time, the lattice ensembles generated by USBSM~\cite{Schaich:2013eba} are being analyzed in search of a light scalar Higgs particle, and we have improved our techniques to extract the effective mass anomalous dimension from the Dirac eigenmode spectrum~\cite{Cheng:2013eu, Cheng:2013bca}.
Although the combination of these complementary studies will shed further light on $N_f = 8$ and its phenomenological viability as the basis of new BSM physics, our results in this work also highlight the importance of comparing studies using different lattice actions, preferably including different fermion formulations, when exploring such unfamiliar and nontrivial systems.

\section*{Acknowledgments} 
We thank Julius Kuti, Daniel N\'ogr\'adi, Zoltan Fodor, Ethan Neil and George Fleming for useful discussions of step scaling and many-flavor physics.
We appreciate ongoing joint efforts with the Lattice Strong Dynamics Collaboration and USBSM community to investigate $N_f = 8$.
This work was supported by the U.S.~Department of Energy (DOE), Office of Science, Office of High Energy Physics, under Award Numbers DE-SC0008669 (DS), DE-SC0009998 (DS, AV) and DE-SC0010005 (AH). 
Numerical calculations were carried out on the HEP-TH and Janus clusters at the University of Colorado, the latter supported by the U.S.~National Science Foundation through Grant No.~CNS-0821794, and on the DOE-funded USQCD facilities at Fermilab.

\section*{Appendix: Data sets}
Tables~\ref{tab:once-smeared} and \ref{tab:twice-smeared} summarize the once- and twice-smeared lattice ensembles considered in this work, respectively.
In all cases we use exactly massless fermions with anti-periodic BCs in all four directions.
For each ensemble specified by the volume and gauge coupling $\be_F$, the tables report the total number of molecular dynamics time units (MDTU) generated with the HMC algorithm, the thermalization cut, and the resulting number of 100-MDTU jackknife blocks used in analyses.
We also list the average plaquette (normalized to 3), to illustrate the roughness of the gauge fields.

\begin{table}[htbp]
  \begin{center}
    \begin{tabular}{ccccc|cccc}
      \hline
      $\be_F$ & MDTU  & Therm.  & Blocks  & Plaq.       & MDTU  & Therm.  & Blocks  & Plaq.       \\
      \hline
              & \multicolumn{4}{c|}{$L = 12$}           & \multicolumn{4}{|c}{$L = 18$}           \\
       5.0    & 3640  & 400     & 32      & 1.19407(8)  & 2440  & 800     & 16      & 1.19396(6)  \\
       5.4    & 3400  & 300     & 31      & 1.29642(9)  & 2480  & 300     & 21      & 1.29634(7)  \\
       5.8    & 4740  & 300     & 44      & 1.39606(7)  & 2160  & 200     & 19      & 1.39607(5)  \\
       6.2    & 6020  & 200     & 58      & 1.48984(7)  & 2960  & 300     & 26      & 1.48976(6)  \\
       6.6    & 6020  & 600     & 54      & 1.57491(6)  & 2200  & 400     & 18      & 1.57488(6)  \\
       7.0    & 6020  & 200     & 58      & 1.65133(6)  & 2630  & 400     & 22      & 1.65134(5)  \\
       7.4    & 6020  & 300     & 57      & 1.71947(6)  & 2070  & 500     & 15      & 1.71950(4)  \\
       7.8    & 6020  & 300     & 57      & 1.78037(6)  & 2900  & 200     & 27      & 1.78039(3)  \\
       8.2    & 6020  & 300     & 57      & 1.83518(5)  & 2140  & 300     & 18      & 1.83515(4)  \\
       9.0    & 5990  & 700     & 52      & 1.92986(5)  & 3500  & 500     & 30      & 1.92979(4)  \\
      10.0    & 5980  & 300     & 56      & 2.02675(4)  & 2870  & 300     & 25      & 2.02674(3)  \\
      11.0    & 5970  & 200     & 57      & 2.10648(5)  & 2300  & 200     & 21      & 2.10637(3)  \\
      \hline
              & \multicolumn{4}{c|}{$L = 16$}           & \multicolumn{4}{|c}{$L = 24$}           \\
       5.0    & 2680  & 200     & 24      & 1.19413(7)  & 1705  & 500     & 12      & 1.19408(6)  \\
       5.4    & 3655  & 300     & 33      & 1.29636(5)  & 1900  & 300     & 16      & 1.29633(3)  \\
       5.8    & 4870  & 300     & 45      & 1.39605(4)  & 2665  & 300     & 23      & 1.39608(4)  \\
       6.2    & 2610  & 200     & 24      & 1.48970(6)  & 1735  & 300     & 14      & 1.48971(3)  \\
       6.6    & 4180  & 300     & 38      & 1.57501(5)  & 2465  & 300     & 21      & 1.57494(3)  \\
       7.0    & 4350  & 300     & 40      & 1.65137(5)  & 2720  & 300     & 24      & 1.65132(3)  \\
       7.4    & 2100  & 100     & 20      & 1.71941(5)  & 2370  & 200     & 21      & 1.71942(3)  \\
       7.8    & 1900  & 300     & 16      & 1.78028(5)  & 2420  & 300     & 21      & 1.78032(2)  \\
       8.2    & 2360  & 200     & 21      & 1.83521(6)  & 2850  & 300     & 25      & 1.83515(3)  \\
       9.0    & 3510  & 200     & 33      & 1.92982(5)  & 2900  & 200     & 27      & 1.92981(3)  \\
      10.0    & 2440  & 300     & 21      & 2.02679(6)  & 3010  & 200     & 28      & 2.02677(2)  \\
      11.0    & 2980  & 200     & 27      & 2.10642(4)  & 1830  & 200     & 16      & 2.10642(2)  \\
      \hline
              & \multicolumn{4}{c|}{$L = 20$}           & \multicolumn{4}{|c}{$L = 30$}           \\
       5.0    &  925  & 300     &  6      & 1.19401(5)  &  760  & 300     &  4      & 1.19402(6)  \\
       5.4    & 1495  & 200     & 12      & 1.29637(6)  & 1070  & 400     &  6      & 1.29634(3)  \\
       5.8    & 1355  & 200     & 11      & 1.39608(6)  & 1430  & 300     & 11      & 1.39606(4)  \\
       6.2    & 1995  & 200     & 17      & 1.48971(6)  &  980  & 300     &  6      & 1.48970(5)  \\
       6.6    & 2195  & 200     & 19      & 1.57495(5)  & 1540  & 200     & 13      & 1.57492(3)  \\
       7.0    & 2030  & 200     & 18      & 1.65130(5)  & 1190  & 300     &  8      & 1.65133(3)  \\
       7.4    & 2420  & 400     & 20      & 1.71946(4)  &  670  & 200     &  4      & 1.71939(4)  \\
       7.8    & 1720  & 200     & 15      & 1.78040(4)  &  680  & 200     &  4      & 1.78042(4)  \\
       8.2    & 1620  & 200     & 14      & 1.83512(5)  &  940  & 200     &  7      & 1.83510(4)  \\
       9.0    & 1435  & 200     & 12      & 1.92980(5)  & 1040  & 300     &  7      & 1.92979(2)  \\
      10.0    & 2270  & 200     & 20      & 2.02678(3)  & 1050  & 200     &  8      & 2.02673(2)  \\
      11.0    & 1940  & 200     & 17      & 2.10641(3)  & 1180  & 500     &  6      & 2.10636(3)  \\
      \hline
    \end{tabular}
  \end{center}
  \caption{\label{tab:once-smeared}Lattice ensembles with one nHYP smearing step.  For each ensemble specified by the volume $L^4$ and gauge coupling $\be_F$, we report the total molecular dynamics time units (MDTU), the thermalization cut, and the resulting number of 100-MDTU jackknife blocks used in analyses.  We also list the average plaquette (normalized to 3), to illustrate the roughness of the gauge fields.}
\end{table}

\begin{table}[htbp]
  \begin{center}
    \begin{tabular}{ccccc|cccc}
      \hline
      $\be_F$ & MDTU  & Therm.  & Blocks  & Plaq.       & MDTU  & Therm.  & Blocks  & Plaq.       \\
      \hline
              & \multicolumn{4}{c|}{$L = 12$}           & \multicolumn{4}{|c}{$L = 18$}           \\
      4.75    & 2940  & 300     & 26      & 1.04188(10) & 2000  & 400     & 16      & 1.04175(6)  \\
      5.0     & 3000  & 200     & 28      & 1.10377(9)  & 2000  & 700     & 13      & 1.10389(5)  \\
      5.25    & 3000  & 200     & 28      & 1.16812(9)  & 2000  & 200     & 18      & 1.16824(6)  \\
      5.5     & 3000  & 400     & 26      & 1.23485(11) & 1978  & 200     & 17      & 1.23490(6)  \\
      5.75    & 3000  & 200     & 28      & 1.30316(11) & 2000  & 300     & 17      & 1.30321(6)  \\
      6.0     & 3000  & 400     & 26      & 1.37229(10) & 1933  & 300     & 16      & 1.37235(4)  \\
      6.25    & 2440  & 400     & 20      & 1.44002(15) & 2000  & 300     & 17      & 1.43999(7)  \\
      6.5     & 2500  & 400     & 21      & 1.50397(13) & 1980  & 200     & 17      & 1.50389(9)  \\
      7.0     & 2500  & 400     & 21      & 1.61717(12) & 1980  & 300     & 16      & 1.61715(8)  \\
      \hline
              & \multicolumn{4}{c|}{$L = 16$}           & \multicolumn{4}{|c}{$L = 24$}           \\
      4.75    & 1910  & 400     & 15      & 1.04191(5)  & 1790  & 200     & 15      & 1.04181(4)  \\
      5.0     & 1990  & 300     & 16      & 1.10379(6)  & 1510  & 400     & 11      & 1.10392(5)  \\
      5.25    & 1800  & 200     & 16      & 1.16829(10) & 1690  & 300     & 13      & 1.16811(5)  \\
      5.5     & 1985  & 300     & 16      & 1.23486(8)  & 1970  & 400     & 15      & 1.23479(4)  \\
      5.75    & 2000  & 200     & 18      & 1.30333(8)  & 2000  & 200     & 18      & 1.30331(3)  \\
      6.0     & 1985  & 300     & 16      & 1.37221(7)  & 1578  & 400     & 11      & 1.37228(3)  \\
      6.25    & 1330  & 200     & 11      & 1.44000(8)  & 1690  & 400     & 12      & 1.43986(4)  \\
      6.5     & 1990  & 400     & 15      & 1.50404(11) & 1550  & 300     & 12      & 1.50392(3)  \\
      7.0     & 1990  & 300     & 16      & 1.61714(7)  & 1310  & 200     & 11      & 1.61711(4)  \\
      \hline
              & \multicolumn{4}{c|}{$L = 20$}           & \multicolumn{4}{|c}{$L = 30$}           \\
      4.75    & 1810  & 300     & 15      & 1.04181(3)  & 1115  & 400     &  7      & 1.04175(5)  \\
      5.0     & 1770  & 400     & 13      & 1.10390(7)  & 1220  & 300     &  9      & 1.10386(3)  \\
      5.25    & 1750  & 200     & 15      & 1.16827(5)  & 1250  & 200     & 10      & 1.16820(4)  \\
      5.5     & 1560  & 400     & 11      & 1.23475(5)  & 1190  & 300     &  8      & 1.23474(4)  \\
      5.75    & 2000  & 300     & 17      & 1.30325(5)  & 2000  & 400     & 16      & 1.30322(2)  \\
      6.0     &  910  & 300     &  6      & 1.37221(9)  & 1498  & 300     & 11      & 1.37225(4)  \\
      6.25    & 1870  & 300     & 15      & 1.43980(4)  & 1600  & 400     & 12      & 1.43986(4)  \\
      6.5     & 2000  & 400     & 16      & 1.50399(5)  & 1740  & 300     & 14      & 1.50401(4)  \\
      7.0     & 2000  & 400     & 16      & 1.61709(5)  & 1710  & 400     & 13      & 1.61708(2)  \\
      \hline
    \end{tabular}
  \end{center}
  \caption{\label{tab:twice-smeared}Lattice ensembles with two nHYP smearing steps, with columns as in \protect\tab{tab:once-smeared}.}
\end{table}
\clearpage 

\bibliographystyle{utphys}
\bibliography{8f_beta}
\end{document}